\documentclass[12pt,a4paper,fleqn]{article}
\usepackage{cite,amsmath,amssymb,bbm,graphicx}
\usepackage{mathrsfs}	
\usepackage{pifont}

\newcommand{\vev}[1]{\langle #1 \rangle}
\newcommand{\D}{\mathrm{d}}

\newcommand{\SU}[1]{\ensuremath{\mathrm{SU}(#1)}}
\newcommand{\U}[1]{\ensuremath{\mathrm{U}(#1)}}
\newcommand{\Z}[1]{\ensuremath{\mathbbm{Z}_{#1}}} 
\newcommand{\W}{\ensuremath{\mathscr{W}}}
\newcommand{\M}{\ensuremath{\mathscr{M}}}

\begin{document}

\thispagestyle{empty}

\vspace*{-2cm}
\begin{flushright}
  DCPT/10/36 \\
  IPPP/10/18 \\
  TUM-HEP 752/10\\
  MPP-2010-25
\end{flushright}
\vspace*{1.2cm}

\begin{center}

 {\Large\bf Approximate $\boldsymbol{R}$-symmetries and the $\boldsymbol{\mu}$ term}\\[12mm]

{\large Felix Br\"ummer$^{\,a}$, Rolf Kappl$^{\,b,c}$, Michael Ratz$^{\,b}$, 
Kai Schmidt-Hoberg$^{\,b}$} 
\\[6mm]

{\it
$^{a}$~Institute for Particle Physics Phenomenology, Durham University,\\ 
Durham DH1 3LE, United Kingdom\\[3mm]
$^{b}$~Physik-Department T30, Technische Universit\"at M\"unchen, \\ 
James-Franck-Stra\ss e, 85748 Garching, Germany\\[3mm]
$^{c}$~Max-Planck-Institut f\"ur Physik (Werner-Heisenberg-Institut),\\
F\"ohringer Ring 6,
80805 M\"unchen, Germany}

\vspace*{12mm}

\begin{abstract}
We discuss the role of approximate $\U1_R$ symmetries for the understanding of
hierarchies in Nature. Such symmetries may explain a suppressed expectation
value of the superpotential and provide us with a solution to the MSSM $\mu$
problem. We present various examples in field theory and string-derived models. 
\end{abstract}

\end{center}

\clearpage

\section{Introduction}

Some of the most fundamental questions in theoretical physics are related to
the observation of huge hierarchies in Nature. According to the traditional
criteria~\cite{'tHooft:1979bh}, such hierarchies are ``natural'' if they are a
consequence of an approximate symmetry. 
Low-energy supersymmetry provides an explanation for the stability, but not the
origin, of the hierarchy between the electroweak scale and the GUT or Planck
scale. To also explain its origin one should find a reason why the  scale of
supersymmetry breakdown is so much smaller than the fundamental scale. In
Minkowski vacua, with gravity-mediated supersymmetry breaking, superpartner 
masses are of the order of the gravitino mass $m_{3/2}$. This in turn is 
essentially given by the vacuum expectation value (VEV) of the superpotential 
$\W$. For supersymmetry to stabilize the electroweak hierarchy, $\vev\W$ should 
be of the order of a TeV, or about $10^{-15}$ in Planck units.

A conventional strategy to obtain vacuum solutions with small supersymmetry 
breaking (and cosmological constant zero) in a given model is to identify  field
configurations in which $\vev\W$ completely vanishes at the perturbative  level.
A suppressed $\vev\W$ and supersymmetry breaking should then be obtained from
non-perturbative effects. This has the appealing feature  that the scale of
supersymmetry breakdown is set by dimensional
transmutation~\cite{Witten:1981nf}, and is thus naturally small. 

Recently it has been argued that the hierarchy between the fundamental and  the
supersymmetry breaking scale might also be due to an approximate $\U1_R$ 
symmetry. Approximate $\U1_R$ symmetries appear naturally in the low-energy 
effective field theories of heterotic string compactifications on orbifolds. 
They may explain a highly suppressed $\vev\W$ in settings in which the typical 
VEVs of the relevant fields are only somewhat below the fundamental
scale~\cite{Kappl:2008ie}.

In any case, even if hierarchically small soft masses are explained, the minimal
supersymmetric extension of the standard model, the MSSM, still suffers from the
so-called $\mu$ problem. That is, the coefficient $\mu$ of the Higgs bilinear
coupling $\mu\,H_u\,H_d$ has to be of the order of the soft masses as well, 
in order to give rise to a reasonable phenomenology. 
There are various proposals to solve this problem \cite{Kim:1983dt,Giudice:1988yz}.

The main point of this study is to show that approximate $R$-symmetries have the
potential to solve both these problems simultaneously. A $\U1_R$ symmetry,
under  which $H_u\,H_d$ is neutral, forbids both the $\mu$ term and any
non-trivial  $\vev\W$. If $\U1_R$ is merely an approximate symmetry, $\vev\W$
can be viewed as an order parameter for $\U1_R$ breaking, setting  consequently
also the size of the $\mu$ term. Hence, a $\mu$ term of the order of the
gravitino mass $m_{3/2}$ can be explained by an approximate $R$-symmetry. We
explore this relation both in the context of field theory and string-derived
MSSM models.

This paper is organized as follows. In section~\ref{sec:SUSYgroundstates} we
recollect some basic facts on supersymmetric ground states and explain how an
(approximate) $R$-symmetry allows us to control the VEV of the superpotential.
In section~\ref{sec:muvsR} we discuss how an $R$-symmetry can relate the $\mu$
term to $m_{3/2}$ in field-theoretic examples with generic superpotentials.
Section~\ref{sec:String} is devoted to a discussion of solutions of the $\mu$
problem in string-derived models. Section~\ref{sec:examples} contains a collection of
example models in which approximate $R$-symmetries lead to suppressed superpotential
VEVs. Finally, section~\ref{sec:Conclusions} contains our conclusions.

\section{Supersymmetric ground states}
\label{sec:SUSYgroundstates}

\subsection{Trivial vs.\ non-trivial field configurations}

Consider a 4D $\mathcal{N}=1$ supergravity theory with chiral superfields
$\Phi_i$ ($1\le i\le M$), which could, for instance, arise as the low-energy 
effective field theory governing the massless spectrum of some string 
compactification. The scalar potential is given by
\begin{equation}\label{eq:Vscalar}
 V~=~
 \mathrm{e}^K\,\left(
 	D_i \W\,D_{\bar\jmath}\overline{\W}\,K^{i\bar\jmath}
 	-3|\W|^2\right)\;.
\end{equation}
Here we ignore possible contributions of $D$-terms and follow the usual notation
conventions: $K$ is the K\"ahler potential, $D_i=\partial_i +(\partial_i K)$ is
the K\"ahler-covariant derivative with respect to the superfield $\Phi_i$, and
$K^{i\bar\jmath}$ is the inverse K\"ahler metric. We work in units where the
Planck scale is set to unity, $M_\mathrm{P}=1$. 

To account for supersymmetry breaking and an (almost) vanishing cosmological
constant, one should find systems in which both terms in the parentheses in 
equation~\eqref{eq:Vscalar} are nonzero and (almost) cancel each other. Furthermore, in 
order to solve the gauge hierarchy problem of the standard model (SM), one is 
lead to consider settings in which $\vev\W\sim10^{-15}$, such that the gravitino 
mass is at the TeV scale.

In string-derived models, calculating the full scalar potential 
is usually impossible for practical purposes. Even in the relatively simple
setting of orbifold compactifications, computing higher-order superpotential
coefficients tends to be fairly cumbersome, and the number of fields entering 
the scalar potential \eqref{eq:Vscalar} is typically of the order of dozens 
or hundreds. For these reasons it is usually not feasible to analytically 
minimize the full scalar potential. However, often one is only interested in 
the low-energy physics appearing from the expansion about some vacuum. A good 
strategy is therefore to make an ansatz for a vacuum configuration, defined
by a set of fields with non-vanishing VEVs, with the 
help of string selection rules and phenomenological considerations. One 
should then show a posteriori that this choice is self-consistent.
The properties of the model which one may deduce from this ansatz should, of 
course, not depend too sensitively on the specific values of the coefficients.

In the traditional approach, one would seek configurations in which
$D_i\W$ and $\W$ vanish at the perturbative level,
leading to a supersymmetric Minkowski vacuum. To be more specific, let us
briefly discuss an example. In \cite{Buchmuller:2006ik} a heterotic orbifold
model (with the precise chiral spectrum of the MSSM) was discussed, in which 
the potential for SM singlets from the second and fourth twisted sectors is
$F$- and $D$-flat to all orders in perturbation theory. This is because any
allowed superpotential coupling involves at least two other fields, such that
$\W$ as well as $\partial_a\mathscr{W}$ vanish (where $a$ runs over the 
SM singlets from the second and fourth twisted sectors). Hence the
scalar potential \eqref{eq:Vscalar} equals zero. That is, the system that 
emerges by setting all other fields to zero has, at the perturbative level, a 
supersymmetric Minkowski ground state, but also a huge moduli space.

Let us now look instead at systems where all fields enter a non-trivial
superpotential. These emerge, for instance, if one switches on other fields in
the above model. An immediate objection against such
systems is that, in order to obtain a small gravitino mass, one has to satisfy
\begin{equation}
 D_i \W~\ll~1 \quad\text{and}\quad \vev \W~\ll~1
\end{equation} 
simultaneously, which constitutes $M+1$ constraints for $M$ fields $\Phi_i$.
In other words, one might expect that for solutions of $D_i \W\ll 1$, which can
be chosen such that the $D$-terms vanish (cf.\ \cite{Luty:1995sd}), one 
would always obtain $\vev\W$ of order one. However, while this argument
certainly applies for completely generic superpotentials, it is not true in the
presence of an approximate $\U1_R$ symmetry. In this case, one obtains
\cite{Kappl:2008ie}
\begin{equation}\label{eq:smallW}
 \vev\W~\sim~{\vev\Phi}^N\;,
\end{equation}
where $\vev\Phi$ denotes the typical VEV scale of the $\Phi_i$ fields and $N$ is
the order at which the approximate $\U1_R$ is explicitly broken in $\W$. If $\vev\Phi$
is slightly suppressed against the fundamental scale
and $N$ is sufficiently large, $\vev\W$ can be
hierarchically small and induce a gravitino mass in the TeV range. In other
words, a mild hierarchy between $\vev\Phi$ and the fundamental scale is
power-law enhanced in the presence of an approximate $R$-symmetry, very similarly
to the well-known Froggatt-Nielsen scheme~\cite{Froggatt:1978nt}. In such
settings one often finds that all $\Phi_i$ fields attain masses, with the
would-be $R$-axion gaining a mass of the order $\W/{\vev\Phi}^2$. In particular,
one obtains point-like vacuum configurations, i.e.\ vacua with dimension zero
moduli space, where a hierarchically small gravitino mass is explained by an
approximate, perturbative symmetry (rather than by non-perturbative effects).

\subsection{Supersymmetric Minkowski vacua as a consequence of $\boldsymbol{\U1_R}$}

Let us review the derivation of equation~\eqref{eq:smallW} in a way which
slightly differs from the one in \cite{Kappl:2008ie}.
Consider first a globally supersymmetric model of chiral superfields which
has an exact $\U1_R$ symmetry. Then, in a supersymmetric vacuum, the
expectation value of the superpotential vanishes: 
since the $R$-charges $r_i$ in each term of $\W$ add
up to $2$, $\W$ is a weighted homogeneous polynomial in the fields satisfying
\begin{equation}\label{whp}
2\,\W~=~\sum_i r_i\,\Phi_i\,\partial_i\W\;.
\end{equation}
In supersymmetric vacua the $\partial_i\W$ vanish, and therefore $\vev \W=0$.

For a model with an exact $\U1_R$ symmetry, a critical point of the
superpotential  necessarily also represents a supersymmetric vacuum in
supergravity: $\partial_i\W=0$ and $\W=0$ imply $D_i\W=0$. This is irrespective
of whether or not the corresponding field configuration represents a physical
vacuum of some globally supersymmetric effective theory, obtained by decoupling
Planck-scale physics.  (In fact, in concrete string-derived models, critical
points of $\W$ often arise from the interplay of higher-dimensional
operators suppressed by powers of $M_\mathrm{P}$, and thus disappear in the
limit $M_{\mathrm{P}}\,\rightarrow\,\infty$.)

Not all supersymmetric vacua in supergravity have $\W=0$. Vanishing $F$-terms imply
$\partial_i\W=-\W\partial_i K$, so that equation~\eqref{whp} becomes
\begin{equation}\label{eq:whpsugra}
 2\,\W~=~-\W\,\sum_i r_i\,\Phi_i\,\partial_iK\;.
\end{equation}
This means that either $\vev\W=0$ as above, or $\sum_i r_i\Phi_i\partial_i K=-2$
in the vacuum. In the following we will only consider supergravity vacua of the
former type, that is with $\vev\W=0$, obtained from critical points of the
superpotential.

In settings with a $\U1_R$ symmetry, the existence of supersymmetric but $R$-breaking vacua
is by itself non-trivial (see, for instance, \cite{Nelson:1993nf}), and will be discussed  
in what follows.

\subsection{A note on supersymmetric vacua with broken $\boldsymbol{R}$-symmetry}
\label{sec:BeyondNelsonSeiberg}

It is sometimes claimed in the literature that spontaneously broken $R$-symmetry
in generic models implies broken supersymmetry (as part of the ``Nelson--Seiberg
theorem'' \cite{Nelson:1993nf}). More precisely, for globally supersymmetric
models\footnote{In this section we work in  global supersymmetry, but the
arguments carry over to supergravity in a straightforward way.} of chiral
superfields with generic superpotentials, the claim is that
\renewcommand{\labelenumi}{(\roman{enumi})}
\begin{enumerate}
 \item if the model admits a supersymmetry-breaking global vacuum, then it
possesses an exact $\U1_R$ symmetry (which may be broken or unbroken in the
vacuum);
 \item if the model possesses an exact $\U1_R$ symmetry and a global vacuum in
which it is spontaneously broken, then the vacuum also breaks supersymmetry
spontaneously.
\end{enumerate}
The cause for concern here is the second statement, since we are relying on
models with supersymmetric vacua in which $R$-symmetry is spontaneously broken.

However, there is a loophole in the Nelson--Seiberg argument, which can be used
to construct counterexamples~\cite{Ray:2007wq}. Let us recall how the
Nelson--Seiberg argument works: assume there are $N$ chiral superfields $\Phi_i$
with $R$-charges $r_i$, normalized such that $\W$ carries $R$-charge $2$. Assume
that $\Phi_N$ breaks $R$-symmetry, so $\vev{\Phi_N}\neq 0$ and $r_N\neq 0$.
Write $\W$ as
\begin{equation}\label{eq:WNS}
 \W
 ~=~
 \Phi_N^{2/r_N}\,f\left(\widetilde{\Phi}_1,\ldots,\widetilde{\Phi}_{N-1}\right)\;,
\end{equation}
where the $\widetilde{\Phi}_j$ are chiral superfields constructed from $\Phi_N$ and
from the other $\Phi_j=\Phi_1,\ldots,\Phi_{N-1}$ as
\begin{equation}
 \widetilde{\Phi}_j~=~\Phi_j\,\Phi_N^{-r_j/r_N}\;.
\end{equation}
The conditions for unbroken supersymmetry are, using that $\Phi_N\neq 0$ and
$r_N\neq 0$ by assumption,
\begin{subequations}\label{eq:UnbrokenSUSYPhiTilde}
\begin{eqnarray}
 0 &= &\frac{\partial \W}{\partial\Phi_j}
~=~\Phi_N^{2/r_N}\,\sum_{k=1}^{N-1}\frac{\partial
f}{\partial\widetilde{\Phi}_k}\frac{\partial
\widetilde{\Phi}_k}{\partial\Phi_j}~=~\Phi_N^{(2-r_j)/r_N}\frac{\partial
f}{\partial\widetilde{\Phi}_j}\qquad(j=1,\ldots,N-1)\;,\\
0&=&\frac{\partial \W}{\partial\Phi_N}
~=~\frac{2}{r_N}\Phi_N^{2/r_N-1}
f+\Phi_N^{2/r_N}\sum_{j=1}^{N-1}\frac{\partial
f}{\partial\widetilde{\Phi}_j}\Phi_j\left(-\frac{r_j}{r_N}\right)\Phi_N^{-r_j/r_N-1}
\end{eqnarray}
\end{subequations}
which is equivalent to 
\begin{equation}\label{SUSYconds}
 f~=~0\quad\text{and}\quad \frac{\partial f}{\partial \widetilde{\Phi}_j}~=~0\;.
\end{equation}
These are $N$ equations in the $N-1$ variables $\widetilde{\Phi}_j$ and thus do not
have a solution if $f$ is a generic function. Therefore there is either no
vacuum, or supersymmetry is spontaneously broken.

The loophole is now that $f$ is not necessarily generic even if $\W$ is. For
example, if $r_N$ is not an integer fraction of $2$, a constant term in $f$ is
not allowed even though it could well be formally compatible with all
symmetries, because it would represent a non-polynomial piece in the
superpotential. Clearly, it is sufficient that $f$ is at least quadratic
in the $\widetilde{\Phi}_i$ fields; then
equations~\eqref{eq:UnbrokenSUSYPhiTilde} are always satisfied at
$\widetilde{\Phi}_i=0$. Such $f$ can be obtained in effective theories where all
massive modes are integrated out and the superpotential $\W$ starts with cubic
terms; examples for such systems include the effective supergravity description
of orbifold compactifications.

A simple example where it is possible to see explicitly how the argument fails
has three chiral superfields $X$, $Y$, $Z$ with the $R$-charges $r_X=3$, $r_Y=1$
and $r_Z=-2$.
The most general superpotential compatible with these charge assignments is
\begin{equation}\label{eq:w123}
 \W~=~Y^2+X\,Y\,Z+X^2\,Z^2+Y^4\,Z+(\text{terms of order 6 and higher})\;.
\end{equation}
We have set all coefficients to $1$, because their values are irrelevant to the
argument as long as they are nonzero. Note that the $R$-charges are
unambiguously fixed by the first four leading-order terms which we have
explicitly written: the quadratic term fixes $r_Y=1$, the quintic term then
fixes $r_Z=-2$, and the other two fix $r_X=3$. 

Neglecting the higher-order terms, the $F$-term equations read
\begin{subequations}
\begin{eqnarray}
F_X&:&\quad Y\,Z+2\,X\,Z^2~=~0\;,\\
F_Y&:&\quad 2\,Y+X\,Z+4\,Y^3Z~=~0\;,\\
F_Z&:&\quad X\,Y+2\,X^2\,Z+Y^4~=~0\;.
\end{eqnarray}
\end{subequations}
They are solved by $Y=Z=0$, with $X$ a flat direction.\footnote{They are also
solved by $X=Y=0$, with $Z$ a flat direction, a case which can be discussed
completely analogously. There are no further solutions.} Along this flat
direction, at any point $X\neq 0$, $R$-symmetry is spontaneously broken while
SUSY is preserved. In fact any vacuum with unbroken supersymmetry but
spontaneously broken $R$-symmetry must have a supersymmetric flat direction,
because the potential must accommodate a Goldstone boson and its complex
partner.

This solution remains unchanged when one takes into account higher-order terms
in $\W$, since these must be at least quadratic in $Z$ and hence give contributions
to the $F$-term equations which are at least linear, vanishing at $Z=0$. 

Following the procedure from the previous section, we can write the
superpotential as
\begin{equation}
 \W~=~X^{2/3}\,\left(\widetilde{Y}^2+\widetilde{Y}\widetilde{Z}
 +\widetilde{Z}^2+\widetilde{Y}^4\widetilde{Z}\right)\;.
\end{equation}
Here $\widetilde{Y}=Y\,X^{-1/3}$ and $\widetilde{Z}=Z\,X^{2/3}$. The term in brackets ($f$
in the above notation) is now not a generic polynomial --- for instance it is
missing a constant term, because this would correspond to a non-polynomial
$X^{2/3}$ term in $\W$.

\subsection{$\boldsymbol{\U1_R}$ breaking at higher order}
\label{sec:HigherOrderRB}

Having seen that also in the presence of a $\U1_R$ symmetry there can be
supersymmetric solutions with nontrivial VEVs, we now turn to discuss the 
impact of higher order, explicit $\U1_R$ breaking superpotential terms.

We are interested in a critical point of the exactly $\U1_R$-symmetric $\W$ in 
which all field VEVs are at least slightly below the Planck scale, 
$\vev{\Phi_i}<1$. Adding explicit $R$-breaking terms of some high order $\gtrsim
N$ in the fields should shift the original expectation values by a small amount
$\epsilon$.

More precisely, in the exactly $\U1_R$ symmetric case there is a flat direction 
in field space, which is the curve parameterizing the VEV of the $R$-axion. Adding explicit 
$R$-breaking terms should lead to an isolated vacuum at a distance $\epsilon$ from 
this curve. If the field expectation values before $R$-breaking were somewhat
small near the new vacuum, it is self-consistent to take $\epsilon$ small as well,
provided $N$ is sufficiently large.

Assuming for simplicity that the original expectation values are all of 
roughly the same size, we can estimate the magnitude of $\epsilon$. Let us 
write $\W$ as the sum of an exactly $R$-symmetric
lower-order part $\W_0$ and an $R$-breaking order-$N$ piece $\Delta \W$,
\begin{equation}
 \W~=~\W_0+\Delta \W\;.
\end{equation}
A supersymmetric vacuum for the $R$-symmetric truncation $\W_0$ at $\Phi^0$ has
$\partial_i\W_0(\Phi^0)=0$, $\W_0(\Phi^0)=0$, and $D_i\W_0(\Phi^0)=0$. If it shifts by 
$\epsilon$ with the full superpotential, and supersymmetry is preserved, then
\begin{eqnarray}
0 
 & = & 
 D_i\W\Bigr|_{\Phi^0+\epsilon}
 ~=~D_i\W\Bigr|_{\Phi^0}+\partial_j D_i\W\Bigr|_{\Phi^0}\epsilon_j
  +\partial_{\bar\jmath} D_i\W\Bigr|_{\Phi^0}\overline{\epsilon}_{\bar\jmath}
  +\mathcal{O}(|\epsilon|^2)\nonumber\\
 & = & D_i \W_0\Bigr|_{\Phi^0}+D_i(\Delta\W)\Bigr|_{\Phi^0}
  +\partial_j D_i\W_0\Bigr|_{\Phi^0}\epsilon_j
  +\partial_{\bar\jmath}D_i\W_0\Bigr|_{\Phi^0}
  \overline{\epsilon}_{\bar\jmath}
  +\mathcal{O}\Bigl(|\epsilon|^2,\;|\Delta \W\,\epsilon|\Bigr)\nonumber\\
 & = & 
  D_i(\Delta\W)\Bigr|_{\Phi^0}+\partial_j \partial_i\W_0\Bigr|_{\Phi^0}\epsilon_j
  +\mathcal{O}\Bigl(|\epsilon|^2,\;|\Delta \W\,\epsilon|\Bigr)\;.
\end{eqnarray}
If $\vev\Phi$ is a typical VEV, then $\vev{\partial_j \partial_i\W_0}\sim\vev\Phi$ for 
superpotentials $\W_0$ which start at cubic order in the fields (as in models
which describe the massless degrees of freedom of some string compactification),
and $\vev{D_i(\Delta W)}\sim{\vev\Phi}^{N-1}$. We obtain the estimate
\begin{equation}
 \epsilon~\sim~\vev{\Phi}^{N-2}\;.
\end{equation}
This shift will cause the terms in $\W_0$ to contribute to a non-vanishing
superpotential expectation value of the order $\vev {\W_0}\sim{\vev\Phi}^N$. The
$R$-breaking term $\Delta \W$ itself will also  directly induce a superpotential
expectation value of the same order.  The resulting 
\mbox{$\vev \W\sim{\vev\Phi}^N$} can be hierarchically small even if  $\vev\Phi$ 
is only moderately small but $N$ is large. 

Let us stress that the above is a very rough estimate whose details can easily
change, depending on the model. However, it should be clear that with
our basic assumptions (sub-Planckian field VEVs and an $R$-breaking term of higher
order) the essential result will remain the same, namely, there is a
hierarchically suppressed $\vev \W$. We will discuss explicit examples in 
section~\ref{sec:examples}.

A higher-order $R$-breaking term turns the formerly Minkowski vacuum into an  
AdS vacuum; since the induced $\vev \W$ is hierarchically small, this translates 
directly into the smallness of the AdS vacuum energy. Additional (possibly
non-perturbative) dynamics should then break supersymmetry to yield
an ``$F$-term uplift'' of the AdS vacuum to a local Minkowski or dS minimum of 
the scalar potential (see e.g.\
\cite{GomezReino:2006dk,Lebedev:2006qq,Brummer:2006dg}). 
Since in such constructions the uplifting sector does not significantly change
$\vev\W$, the gravitino mass in the uplifted model will be given by
the pre-uplift $\vev\W$ and can thus naturally be of the order of a TeV.

Models which do not rely on a separate uplifting sector are also conceivable. 
Consider, for instance, a K\"ahler potential of the 
form~\cite[section~4]{Weinberg:1988cp}
\begin{equation}\label{eq:weinbergK}
 K~=~-3\ln\left(T+\overline{T}-h(C_\alpha,\overline{C}_{\bar\beta})\right)
 +\widetilde{K}(S_n,\overline{S}_{\overline m})\;.
\end{equation}
Here $T$ is the usual K\"ahler modulus, of which the perturbative superpotential
is independent, and $C_\alpha$ and $S_n$ are chiral superfields. The scalar
potential is
\begin{equation}
V~=~\frac{\mathrm{e}^{\widetilde
K}}{(T+\overline{T}-h)^3}\left(\partial_\alpha\W\partial_{\bar\beta}{\overline\W
}\,h^{\alpha\bar\beta}\left(\frac{T+\overline{T}-h}{3}\right)+D_n\W\,D_{
\overline m}\overline\W\,\widetilde K^{n\overline m} \right)\;,
\end{equation}
where $h^{\alpha\bar\beta}$ denotes the matrix inverse of
$\partial_\alpha\partial_{\bar\beta}h$.
Assuming the $T$ modulus to be stabilized, stationary points of $V$ occur at
\begin{equation}\label{weinbergvac}
\partial_\alpha\W~=~D_n\W~=~0\;.
\end{equation}
These stationary points are, in general, not supersymmetric, and they come with 
vanishing vacuum energy (at least at the tree level). Evidently, if $\W$ is 
suppressed at a point where \mbox{$\partial_\alpha\W=\partial_n\W=0$} due to an 
approximate $R$-symmetry, then it will be of the same order in the nearby 
vacuum \eqref{weinbergvac}.

These considerations indicate that one might ultimately even obtain an understanding 
of the smallness of the vacuum energy. Approximate $\U1_R$ symmetries indeed 
allow us to control the vacuum energy, although not to the level needed 
to explain the observed cosmological constant. 

Having outlined the general picture, we now turn to particular classes of models
in which an approximate $R$-symmetry is useful to solve the $\mu$ problem.

\section{Approximate $\boldsymbol{R}$-symmetry and the $\boldsymbol{\mu}$ term}
\label{sec:muvsR}

\subsection{The effective $\boldsymbol{\mu}$ term}
\label{sec:effmu}

In the MSSM there is a single dimensionful parameter at the supersymmetric
level: the Higgsino mass $\mu$. In any model which UV-completes the MSSM at the GUT
scale or at the Planck scale, one would naively expect $\mu$ to be either zero
or of the order of the UV-completion scale. Vanishing $\mu$ is ruled out
experimentally since massless Higgsinos would be in conflict with direct search
limits. A $\mu$ parameter significantly larger than the soft SUSY-breaking Higgs
masses, on the other hand, would require excessive fine-tuning to obtain the
correct $Z$ mass. This in turn would spoil one of the main motivations for
low-energy supersymmetry.

To solve this so-called ``$\mu$ problem'', one should find a mechanism
connecting $\mu$ with the scale of supersymmetry breaking. It should naturally
give a $\mu$ parameter of the order of the gravitino mass $m_{3/2}$, which sets 
the scale for the soft masses. There are some well-known proposals on how to 
obtain a $\mu$ term of the correct size, 
\begin{dingautolist}{"0CA}
 \item from the superpotential \cite{Kim:1983dt}, or
 \item from the K\"ahler potential \cite{Giudice:1988yz}.
\end{dingautolist}

Consider a supersymmetric extension of the MSSM by a number of chiral
superfields $\Phi_i$, which could represent e.g.\ string moduli or some general
hidden sector fields. Let us for the moment ignore MSSM matter fields, since they
will couple to the Higgs fields only by Yukawa terms, and thus be irrelevant for 
the Higgs potential (the coupling between the up-type Higgs and lepton doublets 
can be forbidden by matter parity). The K\"ahler potential and superpotential 
read, in an expansion up to quadratic order in the Higgs fields,
\begin{subequations}\label{eq:KWexpansion}
\begin{eqnarray}
K & = & \mathcal{K} + \mathcal{Y}_u\,|H_u|^2+ \mathcal{Y}_d\,|H_d|^2
+\left(\mathcal{Z}\,H_u\, H_d+\text{h.c.}\right)+\ldots\,,\\
\W & = & \W_0+\widehat{\mu}\,H_u\, H_d+\ldots\,,
\end{eqnarray}
\end{subequations}
where $\mathcal{K}$, $\mathcal{Y}_{u,d}$, $\mathcal{Z}$, $\W_0$ and
$\widehat{\mu}$ are functions of the $\Phi_i$ ($\mathcal{K}$ and
$\mathcal{Y}_{u,d}$ are real, while $\W_0$ and $\widehat{\mu}$ are holomorphic). We
are assuming that none of the $\Phi_i$ carries the same quantum numbers as
either $H_u$ or $H_d$. 

Let us take $\vev{\W_0}$ to be real for convenience. The effective $\mu$ parameter is 
then given by
\begin{equation}\label{muparameter}
\mu ~ = ~ \left(\mathcal{Y}_u\mathcal{Y}_d\right)^{-1/2}
\left(\mathrm{e}^{\frac{\mathcal{K}}{2}}\W_0\,\mathcal{Z}
-\overline F^{\bar\imath}
\frac{\partial\mathcal{Z}}{\partial\overline{\Phi}^{\bar\imath}}
+\mathrm{e}^{\frac{\mathcal{K}}{2}}\widehat{\mu}\right)\;,
\end{equation}
evaluated in the vacuum, with $\overline F^{\bar\imath}=-\mathrm{e}^{K/2}K^{j\bar\imath}D_j \W$. 
On the RHS of equation~\eqref{muparameter}  there are
three contributions to $\mu$ with different origins.  The first term is exactly
the gravitino mass, up to a prefactor 
$\mathcal{Z}/\sqrt{\mathcal{Y}_u\mathcal{Y}_d}$ which is generically of order
one. The second, ``Giudice-Masiero''-type term \cite{Giudice:1988yz} is 
likewise expected to be of the correct order of magnitude since it is
essentially given by $F$-terms. The $\widehat{\mu}$ term, on the other hand, is
a priori not related to SUSY breaking and can give a large contribution of the
order of the fundamental scale.\footnote{ $K$ and $\W$ are defined only up to
K\"ahler--Weyl transformations, with physical quantities depending only on the
invariant $G$-function $G=K+\ln|\W|^2$. This is why it is always possible to
absorb the  $\widehat{\mu}$ term into the K\"ahler potential, working with the
quantities  $\widetilde K=K+f+\bar f$ and $\widetilde\W=\W e^{-f}$,  where
$f=\widehat{\mu}\,H_u\, H_d/\W_0$. This, of course, does not solve the $\mu$ problem but
merely obscures it: in expanding the transformed K\"ahler  potential $\widetilde
K$ as in equation~\eqref{eq:KWexpansion}, $\mathcal{Z}$ will now  pick up a
contribution $\sim \widehat{\mu}/\W_0$. Therefore, if $\W_0$ is small,  $\mathcal{Z}$
grows large and the $\mu$ parameter resulting from  equation~\eqref{muparameter}
remains of the order of the fundamental scale.}

Assume now that the Higgs bilinear $H_u\, H_d$ is a  singlet under all
symmetries dictating the structure of $\W$.

The superpotential is then of the form
\begin{equation}\label{eq:structureofW}
\W~=~\sum_a c_a\,\M_a(\Phi_i)+H_u\, H_d\, \sum_a c_a'\,\M_a(\Phi_i)+\ldots\;,
\end{equation}
i.e.\ $\W_0$ and $\widehat{\mu}$ are given by
\begin{subequations}
\begin{eqnarray}
\W_0 & = & \sum_a c_a\,\M_a(\Phi_i)\,,\label{eq:W0}\\
\widehat{\mu} & = & \sum_a c_a'\,\M_a(\Phi_i)\,.\label{eq:muhat}
\end{eqnarray}
\end{subequations}
Here the $\M_a(\Phi_i)$ are normalized monomials in the $\Phi_i$. They are singlets
under all selection rules except $R$-symmetry. The $c_a$ and $c_a'$ are 
numerical coefficients.

There are now two possibilities to explain why $\widehat{\mu}$ has the correct
size:
\begin{enumerate}
\item The couplings $c_a$ and $c_a'$ may coincide up to a common 
factor $\lambda$ of order one, so $\widehat{\mu}=\lambda\,\W_0$ \cite{Casas:1992mk}.
If, furthermore, $\W_0$ is of the order of a TeV in the vacuum (for instance, 
because of an approximate $R$-symmetry), then the same is true for $\mu$. Such
models appear naturally in heterotic orbifold constructions; they will be discussed 
in section~\ref{sec:String}.\label{casasmunoz}
\item
The couplings $c_a$ and $c_a'$ may be completely uncorrelated, but the superpotential 
may be subject to an approximate $R$-symmetry. As we have shown, in that case $\vev\W$ 
is naturally suppressed. 
In what follows, we will show that, for generic $c_a$, this is due to each
$\M_a$ individually being small in the vacuum (rather than due to large
cancellations between different terms in equation~\eqref{eq:W0}). Therefore also
$\mu$ will be naturally suppressed and of the correct order of magnitude.
\end{enumerate}

\subsection{Models with generic superpotential coefficients}
\label{sec:genericmodels}

Consider a superpotential as in equation~\eqref{eq:structureofW}. Suppose $\W$
is ``generic'' in the following sense: there is a set of
continuous or discrete symmetries under which the fields transform, and only the
terms allowed by these symmetries appear in $\W$, with no fine-tuned
coefficients. We stress that if the superpotential coefficients are correlated
in some manner, or if some of them accidentally vanish, then the arguments to
follow must be modified. This can naturally happen if the model is subject to
certain continuous or discrete non-Abelian symmetries, in which case one is to
apply the subsequent analysis to invariants of such symmetries rather than the
elementary fields.

If there is an exact (for now) $\U1_R$ acting on the $\Phi_i$, with $H_u$ and
$H_d$ uncharged, then $\vev \W=0$ as shown before. At this stage it appears to
be possible that this is achieved by a cancellation of various non-zero terms in
equation~\eqref{eq:W0}. Then $\widehat{\mu}$, by equation~\eqref{eq:muhat},
would give an effective contribution to $\mu$ of the order $M_\mathrm{P}$ (or 
slightly smaller if the $\Phi_i$ expectation values are slightly smaller) since 
the $c_a'$ are arbitrary by assumption.

We will now show that there is in fact no such cancellation for generic 
superpotentials. Generic $\U1_R$-symmetric superpotentials vanish term by 
term, i.e.~monomial by monomial, in their supersymmetric vacua. The proof 
proceeds as follows (see also \cite{Kreuzer:1992bi}): write $\W$ as a sum of 
monomials $\M_a$,
\begin{equation}
 \W~=~\sum_a c_a^0 \M_a
\end{equation}
for some generic set of coefficients $c^0=(c_a^0)$. Suppose that there is a
solution to the $F$-term equations at $\Phi^0=(\Phi_i^0)$:
\begin{equation}
 \frac{\partial \W}{\partial\Phi_i}(\Phi_1^0,\ldots,\Phi_N^0)
 ~=~
 0\quad\forall\,i\;.
\end{equation}
Now for each $c$ in some open neighbourhood $\mathcal{U}$ of $c^0$ in the space of
coefficients, we can construct a corresponding superpotential $\W=\sum c_a \M_a$.
By genericity of $c^0$ we can choose $\mathcal{U}$ such that there exists a
collection of solutions $\left(\Phi_1^0(c),\ldots,\Phi_N^0(c)\right)$ to the
respective $F$-term equations which smoothly depends on $c$. Since each $\W$
vanishes in its supersymmetric vacua, $\W$ vanishes identically on $\mathcal{U}$
when regarded as a function of $c$ via
\begin{equation}
\W(c)~=~\W\,\left(\Phi_1^0(c),\ldots,\Phi_N^0(c);c\right)\;.
\end{equation}
Hence
\begin{equation}
 0~=~\frac{\D \W}{\D c_a}
 ~=~\left[\M_a+\sum_i\frac{\partial\W}{\partial\Phi_i}\frac{\partial\Phi_i}{\partial
 c_a}\right]_{\left(\Phi_1^0(c),\ldots,\Phi_N^0(c)\right)}
 ~=~\M_a\left(\Phi_1^0(c),
\ldots,\Phi_N^0(c)\right)\;,
\end{equation}
which proves the assertion.

To return to the $\mu$ problem, we have shown that, as long as the
superpotential is generic and $\U1_R$-symmetric, all $\M_a$ in
equation~\eqref{eq:structureofW} vanish in a supersymmetric vacuum. This implies
that the potentially dangerous contribution $\widehat{\mu}$ to the $\mu$ parameter in 
equation~\eqref{muparameter} also vanishes. Introducing a
higher-order $R$-breaking term, which induces a superpotential expectation value
$\vev \W\sim m_{3/2}$, will also induce a $\widehat{\mu}$ term of the same order of 
magnitude, and thus give a $\mu$ parameter of the correct size.

\section[$\mu$ term in string-derived models]{The $\boldsymbol{\mu}$ term in string-derived models}
\label{sec:String}

Recently explicit string-derived models with approximate $R$-symmetries have
been obtained in which the combination $H_u\,H_d$ is a singlet w.r.t.\ all 
symmetries~\cite{Lebedev:2007hv}. By the arguments of the previous section,
$\mu$ would then be of the order of the gravitino mass if the superpotential
couplings were generic and uncorrelated.

On the other hand, it is not really clear if arguments based on `genericity' 
can be applied to string models, where coupling strengths are calculable and 
satisfy highly non-trivial consistency criteria. Yet the $\mu$ problem is 
solved in certain settings because the superpotential exhibits a non-trivial 
structure, which relates the coupling coefficients. 

The question of $\mu$ terms in string-derived models has been analyzed in the
past in the context of orbifold compactifications of the heterotic string, and
it has been found that, indeed, scenarios incorporating a solution to the $\mu$
problem exist (although concrete models have not been presented). According to our classification in
section~\ref{sec:effmu}, these scenarios fall into two classes:
\begin{dingautolist}{"0CA}
 \item $\mu$ from the superpotential \cite{Casas:1992mk};
 \item $\mu$ from the K\"ahler potential
 \cite{Antoniadis:1994hg,Brignole:1996xb}.
\end{dingautolist}
In what follows we will show that both scenarios are related, at least in
certain explicit string-derived models.

Both \ding{"0CA} and \ding{"0CB} require that the pair $H_u\,H_d$ be vector-like
(not only w.r.t.\ the standard model gauge group
$G_\mathrm{SM}=\SU3_\mathrm{C}\times\SU2_\mathrm{L}\times\U1_Y$, but also
w.r.t.\ all other symmetries). Further, for the second scenario (\ding{"0CB})
to work, the Higgs pair has to come from the untwisted sector, specifically from
an orbifold plane with \Z2 symmetry. In what follows we will briefly review both
scenarios and show that both emerge automatically and simultaneously from
a subset of the MSSM models of the heterotic Mini-Landscape
\cite{Lebedev:2006kn,Lebedev:2007hv}.

Scenario \ding{"0CA} requires the superpotential to be of the form
mentioned in point (i) of section \ref{sec:effmu} \cite{Casas:1992mk}
\begin{equation}\label{eq:CasasMunozW}
 \mathscr{W}~=~\mathscr{W}_0+\lambda\,\mathscr{W}_0\,H_u\,H_d\;,
\end{equation}
where $\mathscr{W}_0$ denotes the superpotential of the hidden sector, which is
responsible for supersymmetry breakdown. It is clear that, once $\mathscr{W}_0$
acquires a VEV, an effective $\mu$ term 
$\widehat{\mu}=\lambda\,\langle\mathscr{W}_0\rangle$ is induced. Since in vacua with
vanishing cosmological constant $\langle\mathscr{W}_0\rangle\sim m_{3/2}$, it
is automatically of the right size.

Scenario \ding{"0CB} relies on the special form of the K\"ahler potential for
untwisted matter fields
\cite{Cvetic:1988yw,Antoniadis:1994hg,Brignole:1996xb,Brignole:1997dp}, 
\begin{equation}\label{eq:KaehlerHiggs}
 K~=~
 -\ln\left[
 	\left(T+\overline{T}\right)\,\left(Z+\overline{Z}\right)
 	-\left(H_u+\overline{H_d}\right)\,\left(H_d+\overline{H_u}\right)
 \right] \;.
\end{equation}
Here, $T$ denotes the K\"ahler modulus of an orbifold plane with a \Z2
symmetry and $Z$ the corresponding complex structure modulus. Expanding $K$ in
the $H_u$ and $H_d$ fields leads to a coupling (essentially
$\overline{T}\,H_u\,H_d+\text{c.c.}$) as is required for the Giudice-Masiero
mechanism to work. Assuming a (dominant) VEV of the $F$-component of $T$ leads
to an effective $\mu$ term just like in the Giudice-Masiero mechanism. It is of
the correct size provided $\widehat{\mu}$ in equation~\eqref{muparameter} is
absent, or at most of the order of $m_{3/2}$.

Before we consider specific models, we note that any Higgs pair coming from
the untwisted sector in an orbifold plane with \Z2 symmetry, the superpotential
has automatically the structure of equation~\eqref{eq:CasasMunozW}. More specifically, we
will show that then the Higgs pair $H_u$ and $H_d$ is neutral w.r.t.\ the
selection rules, i.e.\ whenever a monomial
\begin{equation}\label{eq:monomial}
 \mathscr{M}~=~\prod_i \Phi_i
\end{equation}
denotes a superpotential term, i.e.\ is allowed by the selection
rules~\cite{Dixon:1986qv}, also the term
\begin{equation}\label{eq:monomialHuHd}
 \mathscr{M}\,H_u\,H_d~=~H_u\,H_d\,\prod_i \Phi_i
\end{equation}
is allowed. This has been noted to be the case for the Higgs field of a heterotic
benchmark model of \cite{Lebedev:2007hv}. The argument turns out to be generally
valid. Because of the \Z2 projection, the pair is vector-like w.r.t.\ all gauge
factors. Further, the $\Z2^R$-charges are $(0,0,-1)$ for both, and the
corresponding discrete $R$-symmetry says that $R$ charges should add to $-1\mod
2$ in the third component. Hence the pair $H_u\,H_d$ has $R$-charges which are
equivalent to $(0,0,0)$, i.e.\ $H_u\,H_d$ is neutral w.r.t.\ the $R$ charges.
Moreover, these fields come from the untwisted sector and correspond therefore
to the space group element $(\mathbbm{1},0)$, i.e.\ they are neutral under the
discrete symmetries representing the space group rule. Altogether we have found
that the pair $H_u\,H_d$ is always neutral w.r.t.\ all selection rules, not only
in the $\Z6$-II orbifold. For instance, the $\Z2\times\Z2$ model
presented in \cite{Blaszczyk:2009in} exhibits this structure as well. However,
this argument does not tell us that the coefficients of the terms coincide. That
is, at $H_u\,H_d=0$ the superpotential is as in equation~\eqref{eq:W0},
\begin{equation}\label{eq:superpotential}
 \mathscr{W}~=~\sum c_a\,\mathscr{M}_a\;,
\end{equation}
where the $\mathscr{M}_a$ are monomials of type \eqref{eq:monomial}, while the
terms multiplying $H_u\,H_d$ are as in equation~\eqref{eq:muhat},
\begin{equation}\label{eq:huhd}
 \sum c_a'\,\mathscr{M}_a\;.
\end{equation}
Now, if $\langle\mathscr{W}\rangle$ is small due to the cancellation between
various $\langle\mathscr{M}_a\rangle$, the effective $\mu$ term $\widehat{\mu}=\sum
c_a'\,\langle\mathscr{M}_a\rangle$ is not necessarily small, unless the $c_a$
and $c_a'$ are proportional to each other.

There is a simple way to see that they are indeed proportional. Our starting
point is the K\"ahler potential \eqref{eq:KaehlerHiggs}. Let us expand it, 
\begin{eqnarray}
 K
 & = &
 -\ln\left[
 	\left(T+\overline{T}\right)\,\left(Z+\overline{Z}\right)
 	-\left(H_u+\overline{H_d}\right)\,\left(H_d+\overline{H_u}\right)
 \right]
 \nonumber\\
 & \simeq &
 -\ln\left[
 	\left(T+\overline{T}\right)\,\left(Z+\overline{Z}\right)
 \right]
 + \frac{1}{\left(T+\overline{T}\right)\,\left(Z+\overline{Z}\right)}
  	\left[|H_u|^2+|H_d|^2+(H_u H_d + \text{c.c.})\right]
 \nonumber\\	
 & = &
 -\ln\left[
 	\left(T+\overline{T}\right)\,\left(Z+\overline{Z}\right)
 \right]
 + \left[|\widehat{H}_u|^2+|\widehat{H}_d|^2+(\widehat{H}_u \widehat{H}_d + \text{c.c.})\right]
  \;.
\end{eqnarray}
In the last step we switched to canonically normalized fields
$\widehat{H}_{u,d}$. We see that the constants $\mathcal{Y}_u$, $\mathcal{Y}_d$
and $\mathcal{Z}$ in \eqref{eq:KWexpansion} all coincide. We further make the
assumption that $T$ and $Z$ are already stabilized (see~\cite{Dundee:2010sb} for
a recent discussion; alternative ideas for moduli stabilization are sketched in
section~\ref{sec:string-inspired_example}).

The structure of the K\"ahler potential \eqref{eq:KaehlerHiggs} is enforced by 
gauge invariance in higher dimensions (cf.~\cite{Choi:2003kq,Brummer:2009ug}). The main point
is that $H_u$ and $H_d$ can be viewed as extra components of gauge fields in a
6D (Burdman-Nomura-type~\cite{Burdman:2002se}) orbifold GUT limit of the setting
(see \cite{Hosteins:2009xk} for a recent discussion). Gauge transformations
along the generators which correspond to $H_u$ and $H_d$ mix $H_u$ with
$\overline{H}_d$; therefore the K\"ahler potential can, at the gauge symmetric
level, only depend on the absolute square of $H_u+\overline{H}_d$.  6D gauge
invariance is broken by the boundary conditions of the orbifold
compactification. Therefore, there are the usual logarithmic corrections to
$\mathcal{Y}_u$, $\mathcal{Y}_d$ and $\mathcal{Z}$ below the compactification
scale. Moreover, there are further logarithmic corrections coming from the fact
that at the massless level states sitting at the fixed points do not furnish
complete \SU6 representations.

6D gauge invariance has further important implications. Because of the above
arguments,  the gauge-invariant superpotential is 
\begin{equation}
 \mathscr{W}~=~\text{independent of the monomial}~\widehat{H}_u
 \widehat{H}_d\;.
\end{equation}
Then this setting is, in leading order in $\widehat{H}_u\widehat{H}_d$,
equivalent to a setting with
\begin{subequations}
\begin{eqnarray}
  \widetilde K & = &-\ln\left[
 	\left(T+\overline{T}\right)\,\left(Z+\overline{Z}\right)
 \right]
 + \left[|\widehat{H}_u|^2+|\widehat{H}_d|^2\right] \;,
 \\
 \widetilde{\mathscr{W}} & = & \exp(\widehat{H}_u\,\widehat{H}_d)\,\W\;,
 \label{eq:factorizedsuperpotential}
\end{eqnarray}
\end{subequations}
where the proportionality discussed around equations \eqref{eq:superpotential}
and \eqref{eq:huhd} is obvious.
This statement can easily be verified by looking at the K\"ahler
function $G$ (compare the discussion in section \ref{sec:effmu}). In leading 
order in $\widehat{H}_u\,\widehat{H}_d$ we have
\begin{equation}
 \exp\left(K + \ln|\W|^2\right) 
 ~=~ 
 \exp\left(\widetilde K + \ln|\widetilde{\mathscr{W}}|^2\right)\;.
\end{equation}

Let us also remark that the factorized structure
\eqref{eq:factorizedsuperpotential} can also be obtained by a naive field-theoretic
calculation of the coupling constants.  As $H_u$ and $H_d$ come from the
untwisted sector or, in other words, are extra components of the gauge fields in
ten dimensions, they are bulk fields in the ten-dimensional theory. Since the
product $H_u\,H_d$ is completely neutral under all symmetries, the profile of 
$H_u\,H_d$ is flat; this remains true if one includes effects that distort the
profiles of charged fields, such as localized Fayet-Iliopoulos terms (cf.\ the
discussion in~\cite{Lee:2003mc}). In the naive field-theoretic approach
calculating the couplings amounts to computing the overlap (for canonically
normalized $H_u$ and $H_d$)
\begin{equation}\label{eq:overlap}
 \int\!\D^6y\,\left[H_u(y)\,H_d(y)\right]^n\,\mathscr{M}_a\bigl(\Phi_i(y)\bigr)
\end{equation}
in the internal six dimensions parametrized by $y_M$. Due to the flatness of the
profile of $H_u\,H_d$, the integrals \eqref{eq:overlap} coincide for all $n\ge0$
(up to a constant) with $\int\!\D^6y\,\mathscr{M}_a\bigl(\Phi_i(y)\bigr)$, which
yields the coefficient $c_a$ of the monomial $\mathscr{M}_a$. Including
combinatorial factors again leads to the exponential structure in 
equation~\eqref{eq:factorizedsuperpotential}.

Altogether we see that the holomorphic coupling $H_u\,H_d\,\mathscr{W}$ comes, in
the above supergravity formulation~\cite{Antoniadis:1994hg},  from the K\"ahler
potential. For small $H_u\,H_d$, we can write it into the
superpotential.\footnote{This is not in contradiction with the stringy
calculation of allowed superpotential couplings, which are the basis of the
statements around equations~\eqref{eq:monomial} and \eqref{eq:monomialHuHd}.
This analysis, which is also extensively used in heterotic model building  (as for instance in
\cite{Lebedev:2007hv}), shows only if a holomorphic correlator between
certain fields exists (or not).} In the supergravity formulation it is easy to
show that $\mu\propto\langle\mathscr{W}\rangle$, which requires additional
assumptions in the ``$\mu$ from $\mathscr{W}$'' approach~\cite{Casas:1992mk}.

We also comment that, if $T$ and/or $Z$ attain non-trivial $F$-term VEVs, there
is, as discussed below equation~\eqref{eq:KaehlerHiggs}, an additional,
non-holomorphic contribution to $\mu$
\cite{Antoniadis:1994hg,Brignole:1996xb,Brignole:1997dp}. That is, in the
string-derived MSSM models both the Kim-Nilles \cite{Kim:1983dt} and the
Giudice-Masiero \cite{Giudice:1988yz} mechanisms can be at work, where the
former is always there and the second might or might not contribute. In
particular, no $F$-term expectation value of the `radion' $T$ is required in
order to generate the $\mu$ term.

In conclusion, although in string theory couplings are not `generic' but highly
constrained by consistency and, in particular, calculable, there exist simple
settings in which the $\mu$ problem is solved~\cite{Antoniadis:1994hg} in the
sense that $\mu\sim m_{3/2}$. These settings are incorporated in very promising
orbifold models
\cite{Buchmuller:2005jr,Buchmuller:2006ik,Lebedev:2006kn,Lebedev:2007hv,Blaszczyk:2009in},
which also exhibit the appropriate approximate $R$-symmetries allowing us to
understand why $\vev\W$ is small in the first place. These approximate
symmetries are a consequence of high power discrete $R$-symmetries, reflecting
the discrete rotational symmetries of compact space. In what follows, we will
illustrate the suppression of $\vev\W$ due to approximate $R$-symmetries in
various examples.

\section{Examples}
\label{sec:examples}

\subsection{A simple example}
\label{sec:SimpleExample}

A very simple example to illustrate the mechanism is given by a model with
only two superfields $X$ and $Y$, where $X$ carries $R$-charge 2 and $Y$ has
zero $R$-charge.
At the $\U1_R$-symmetric level, the most general superpotential is
\begin{equation}
 \W~=~X\,f(Y)\;,
\end{equation}
where $f$ is an arbitrary function. Suppose that $\W$ can be written as
\begin{equation}\label{eq:XYW}
 \W~=~X\left(\lambda\, Y^2+\frac{1}{M}\,Y^3+\ldots\right)
\end{equation}
with higher-order terms omitted. The $F$-term equations have a non-trivial solution at
\begin{equation}\label{eq:XYvac} 
 \vev{X}~=~0\;,\qquad \vev{Y}~=~-\lambda\, M\;.
\end{equation}
We will make the assumption that $\vev{Y}$ is somewhat smaller than the fundamental
scale. \eqref{eq:XYvac} is clearly a local minimum. Expanding around $\vev Y$, 
i.e.\ replacing $Y$ by $\vev Y + \delta Y$, leads to
\begin{equation}
\W~=~m\, X\,\delta Y+\ldots
\end{equation}
with $m = -2\, \lambda^2\, M$. The important point is that $m$ is related to the 
fundamental scale and does not know of supersymmetry breakdown.

Now add a $\U1_R$ violating term $Y^N$, i.e.\
\begin{equation}\label{eq:XYWRbreaking}
\W~=~X\left(\lambda\, Y^2+\frac{1}{M}\,Y+\ldots\right)+\kappa\, Y^N\;.
\end{equation}
Then the above minimum undergoes a small shift, but remains a minimum if
$\kappa$ is sufficiently small and/or $N$ is sufficiently large. The expectation
value of the superpotential is of the order $\kappa(\lambda M)^N$. This
eventually sets the scale  for the gravitino mass,
\begin{equation}
 m_{3/2}~\sim~\vev\W~\sim~\kappa\,(\lambda\, M)^N\;.
\end{equation}
The masses of the fluctuations around the slightly shifted minimum are still of
order $m$, i.e.\ can be much larger than $m_{3/2}$. The gravitino mass can be 
arbitrarily small if $\vev Y$ is slightly suppressed and $N$ is sufficiently
large.

\subsection{A ``string-inspired'' example}
\label{sec:string-inspired_example}

Consider now a supersymmetric field theory with matter fields $X$, $Y$ and $Z$
where $X$ and $Y$ have $R$-charge 2 while $Z$ has $R$-charge 0. Our
superpotential is now
\begin{equation}
 \mathscr{W}~=~X\,\left(\lambda_1(T)\,Z^2+\frac{a_1}{M}Z^3+\dots\right)
 +Y\,\left(\lambda_2(T)\,Z^2+\frac{a_2}{M}Z^3+\dots\right)\;.
\end{equation}
Again, the higher order terms ``$\dots$'' will be ignored. Here we have taken
into account a possible dependence of the couplings on the moduli, represented
by $T$. That is, the true field content of our setting is $\{X,Y,Z,T\}$.
Consider now the analogue of the vacuum configuration in
example~\ref{sec:SimpleExample}. We seek solutions to the $F$ equations at $X=Y=0$
with non-trivial $Z$. While the $F$-term equations for $Z$ and $T$ are trivially
satisfied, the other two equations yield
\begin{subequations}\label{eq:Feqs2}
\begin{eqnarray}
 F_X & : & \lambda_1(T)\,Z^2+\frac{a_1}{M}Z^3 ~=~0\;,\\
 F_Y & : & \lambda_2(T)\,Z^2+\frac{a_2}{M}Z^3 ~=~0\;.
\end{eqnarray}
\end{subequations}
This constitutes two equations for the fields $Z$ and $T$. $T$ will be fixed by
\begin{equation}\label{eq:lambdarelations}
 \frac{\lambda_1(\langle T\rangle)}{\lambda_2(\langle T\rangle)}~=~\frac{a_1}{a_2}\;.
\end{equation}
(For instance, if $\lambda_i(T)~=~\mathrm{e}^{-b_i\,T}$, $T$ will be fixed at
$\vev T=\ln(a_1/a_2)/(b_2-b_1)$.) Plugging this back allows us to solve for $Z$,
$\vev Z =-\lambda_1(\langle T\rangle)\,M/a_1$. Again, all fields are
fixed. As long as the $\lambda_i(\langle T\rangle)$ are not too small, the
masses of the fields are not too far below the fundamental scale. The important
lesson here is that moduli, governing the couplings between ``matter fields'',
can be fixed if the $F$-term equations for the matter fields alone are
``overconstraining''. Of course, we have to make the assumption that the
$\lambda_i$ and $a_i$ are such that they admit solutions of 
equation~\eqref{eq:lambdarelations} with $\lambda_i(\langle T\rangle)$ 
somewhat below 1.
We will yet have to see if this mechanism allows us to stabilize the $T$-moduli
in honest string-derived models. As before, adding higher-order $\U1_R$-breaking
terms will result in a suppressed expectation value of $\W$.

In this and the previous examples, there is no symmetry principle enforcing the
structure of the $\U1_R$-symmetric superpotential \eqref{eq:XYW}, nor the
specific structure \eqref{eq:XYWRbreaking} of the $R$-breaking term. Arguing on
the purely field-theoretic level, there is no reason why there should not be
linear terms in $X$, $Y$ or $Z$, or why the leading $R$-breaking term should be
of some high order $N$. (Once such features are imposed, they are however robust
under radiative corrections because of the non-renormalization theorem.) In what
follows, we will present a class of generic, albeit more complicated models, in
which the absence of these terms is enforced by symmetries.

\subsection{A generic example in field theory}

Consider three fields $X$, $Y$, $Z$ which are charged under a 
$\mathbb{Z}_9 \times \mathbb{Z}_4$ symmetry with charges:
\begin{center}
\begin{tabular}{|l||c|c|c|}
\hline
field  & $X$ & $Y$ & $Z$\\ \hline
$\mathbb{Z}_9$-charge & $1$ & $5$ & $8$ \\ 
$\mathbb{Z}_4$-charge & $0$ & $3$ & $3$ \\
\hline
\end{tabular}
\end{center}
The corresponding superpotential reads at order 9
\begin{equation}
\W ~=~ \lambda_5\, X\, Y^2\, Z^2 
+ \lambda_8\, X^4 \,Y^3\, Z + \lambda'_8\, X^4\, Z^4 + \lambda_9\, X^9 \;.
\end{equation}
If truncated at order 8, $\W$ exhibits an accidental $\U1_R$ symmetry with charges
$q^i_R=\left(0,\tfrac{1}{2},\tfrac{1}{2}\right)$; the identification of such
symmetries is described in appendix~\ref{app:AccidentalRSymmetries}.  
$\U1_R$ is explicitly broken at order nine.
One solution to the global $F$ equations is
\begin{subequations}
\begin{eqnarray}
\vev X & = & - \frac{(-2)^{2/9}\lambda_5^{1/3}}{3^{1/3}\lambda_8^{2/9}
\lambda_8'^{1/9}}\;, \\
\vev Y & = & - \frac{(-2)^{11/18}\lambda_5^{5/12}}{3^{5/12}\lambda_8^{11/18}
\lambda_8'^{1/18}} \lambda_9^{1/4}\;, \\
\vev Z & = &  \frac{(-2)^{5/18}\lambda_5^{5/12}}{3^{5/12}\lambda_8^{5/18} \lambda_8'^{7/18}} \lambda_9^{1/4}
\;,
\end{eqnarray}
\end{subequations}
resulting in 
\begin{equation}
\vev \W~=~ - \frac{4 \lambda_5^3}{27 \lambda_8^2 \lambda'_8} \lambda_9 \;.
\end{equation}
If the VEVs are somewhat small, the VEV of the superpotential will be suppressed. All 
masses turn out to be non-vanishing.

In the $\U1_R$-symmetric limit $\lambda_9 \rightarrow 0$, $\vev X$ 
remains finite while $\vev Y$ and $\vev Z$ go to zero. Therefore $\vev\W$ 
vanishes term by term in this limit, as it should by the arguments of 
section~\ref{sec:genericmodels}. It turns out that $Y$ and $Z$ in fact
become massless as $\lambda_9 \rightarrow 0$, with their expectation values
determined by higher-power terms in the scalar potential.

\subsection{Another generic example in field theory}

We now show that it is possible to construct a generic model satisfying the
following requirements: for the $\U1_R$-symmetric truncation there is a 
supersymmetric vacuum at sub-Planckian expectation values. This vacuum breaks 
$\U1_R$ spontaneously. Therefore the Nelson--Seiberg argument must be 
circumvented, and the vacuum in the $\U1_R$ symmetric 
truncation will possess at least one flat direction. It turns out that, in
models where this flat direction includes the $R$-symmetric point where $\U1_R$
is not spontaneously broken, higher-order $R$-breaking terms tend to stabilize
the flat direction at this point. But if the VEVs of all $R$-charged fields 
vanish, higher-order terms will not induce a non-vanishing $\vev\W$. Therefore 
we focus on settings with a flat direction along which $\U1_R$ is spontaneously 
broken everywhere.

The following class of examples satisfies these criteria. It comprises three chiral
superfields $X$, $Y$ and $Z$ with the following $R$-charges:
\begin{center}
\begin{tabular}{|l||c|c|c|}
\hline
field  & $X$ & $Y$ & $Z$\\ \hline
$R$-charge & $2$ & $3$ & $-3$\\ \hline
\end{tabular}
\end{center}
The most general superpotential is
\begin{equation}\label{eq:mostgeneralW}
\W~=~X\,f\bigl(Y\,Z, X^3\,Z^2\bigr)\;.
\end{equation}
There are supersymmetric vacua in configurations with $X=0$ and $Y\,Z=\alpha$, 
where $\alpha$ is determined by the condition $f(\alpha,0)=0$.

The superpotential up to order $10$ in the fields can be written as
\begin{equation}
\W~=~X\,P(YZ)+X^4 Z^2\,Q(YZ)+\dots\;,
\end{equation}
where terms of order 11 and higher have been omitted, $P$ is a quartic
polynomial, and $Q$ is a quadratic polynomial. Supersymmetric vacua appear at
$X=0$ and at the zeros of $P$. Assume that $P$ has an isolated zero at
$YZ=\alpha$ with $\alpha\lesssim 1$ real. We will eventually show that it is
self-consistent to neglect the higher-order terms, provided that $\alpha$ is
somewhat small (and that their coefficients are not too large).

We now add higher-order $R$-breaking terms. To justify their absence at lower
orders, we take the $R$-symmetry not to be $\U1$ but to be given by a discrete
subgroup. Let us consider $\mathbb{Z}_{16}$ for illustration, demanding $R\equiv
2$ mod $16$ for each superpotential term. This allows for several more terms in
$\W$,
\begin{eqnarray}\label{DeltaW}
\lefteqn{\Delta\W~=~ \lambda_1\,Y^6+\lambda_2\, Y^7 Z + \lambda_3\, Y^8 Z^2 +
\lambda_4\,Z^{10}} \nonumber \\ 
&& {}+\beta_1\,X^9 + \beta_2\,X^6\, Y^2 +\beta_3\,X^6\, Y^3\, Z+ 
\beta_4\,X^3\, Y^4+\beta_5\,X^3 \,Y^5 Z + \beta_6\,X^2\, Z^6
+\beta_7\, X^2\, Y Z^7\nonumber \\
&& {}+\dots\;.
\end{eqnarray}
The terms in the first line stabilize the flat direction (by contrast, the terms
in the second line do not contribute to the $F$-terms when evaluated at $X=0$).
We find that  there is a supersymmetric vacuum at
\begin{subequations}
\begin{eqnarray}
\vev{X}&\approx&-\frac{2\,(3\lambda_1)^{5/8}\,(5\lambda_4)^{3/8}}{P'(\alpha)}\,\alpha^{11/4}\;,  \\
\vev{Y}&\approx& \left(\frac{5\lambda_4}{3\lambda_1}\right)^{1/16}\alpha^{5/8}\;, \\
\vev{Z}&\approx& \left(\frac{3\lambda_1}{5\lambda_4}\right)^{1/16}\alpha^{3/8}\;.
\end{eqnarray}
\end{subequations}
The expressions for $Y$ and $Z$ are obtained from the equations of motion
of the $\U1_R$-symmetric theory, up to the VEV along the flat direction. This 
VEV is then calculated by taking into account the higher $\U1_R$-breaking 
terms. The expression for $X$ results from re-substituting 
the $Y$ and $Z$ VEVs into the $F$-term equations of $Y$ and $Z$ and
solving them to leading order. For small $\alpha$ this is a
self-consistent procedure, leading to exact values asymptotically as $\alpha$ 
approaches zero.

The vacuum expectation value of $\W$ behaves like $\alpha^{15/4}$, with the
dominant contributions coming from the $\lambda_1$ and $\lambda_4$ terms in
$\Delta\W$. Clearly no excessive fine-tuning of coefficients is required to
obtain smallish expectation values and suppressed $\vev\W$ (a mildly suppressed
$\alpha$, as we have assumed, is sufficient). At small expectation values, it is
in particular safe to neglect higher-order terms in $\W$.

The superpotential VEV is not suppressed by a very large power here, but it is
straightforward to extend this model to higher suppression of
$\vev\W$, by imposing a $\mathbb{Z}_N$ $R$-symmetry with $N$ sufficiently large.
However, then $\W$ becomes rather cumbersome to write down explicitly. 
Following the procedure described above, one can determine the 
$\alpha$-dependence of $Y$ and $Z$ to leading order in an asymptotic expansion, 
to infer the behaviour of $\vev\W$ at small $\alpha$. The result for some 
select choices of $N$ is:
\begin{center}
\begin{tabular}{|l||c|c|c|c|c|c|c|c|c|c|}
\hline
 $N$ & 10 & 13 & 14 & 16 & 17 & 19 & 20 & 22 & 25 & 28 \\ \hline 
$\vev\W\sim$ & $\alpha^{12/5}$ & $\alpha^{40/13}$& $\alpha^{20/7}$& $\alpha^{15/4}$& $\alpha^{60/17}$& $\alpha^{84/19}$& $\alpha^{21/5}$& $\alpha^{56/11}$& $\alpha^{144/25}$& $\alpha^{45/7}$
\\ \hline
\end{tabular}
\end{center}
The determination of the exponents of $\alpha$ works as follows: the
$\U1_R$-symmetric potential has a flat direction along the hyperbola
$Y\,Z=\alpha$. To stabilize it one needs two $R$-breaking terms in the
superpotential, of the form $\Delta\W\supset Y^p+Z^q$. A $Y^p$ term prevents a
runaway towards $Z=0$ and $Y\rightarrow\infty$; likewise, a $Z^q$ term prevents
a runaway towards $Y=0$ and $Z\rightarrow\infty$. Therefore, the lowest powers
$p$ and $q$ allowed by the discrete $R$-symmetry will essentially determine the
leading $\alpha$-dependence of $Y$ and $Z$, and also of $\W$. In other words,
given $N$, one just has to determine $p$ and $q$ in order to figure out the
exponents of $\alpha$ in the table.
We have also checked this behaviour numerically for several of the above $N$,
using generic superpotentials with random real coefficients of order one.

Other variations of this model can be found by choosing different $R$-charge
assignments for $Y$ and $Z$. For instance, with a $\mathbb{Z}_{16}$ $R$-symmetry
and the $R$-charges

\begin{center}
\begin{tabular}{|l||c|c|c|}
\hline
field  & $X$ & $Y$ & $Z$\\ \hline
$R$-charge & $2$ & $3/4$ & $-3/4$\\ \hline
\end{tabular}
\end{center}
the operators most relevant for stabilizing the $R$-flat direction are $Y^{24}$
and $Z^{40}$, leading to a superpotential expectation value which scales as
$\vev\W\sim\alpha^{15}$. For a mildly suppressed $\alpha$ (or equivalently
mildly suppressed $Y$ and $Z$ VEVs), a huge hierarchy is generated.

So far the discussion has been on the globally supersymmetric level, in the
sense that we are identifying vacua as points where all derivatives of the
superpotential vanish. Using ordinary derivatives rather than K\"ahler-covariant
derivatives may seem not well justified, since some of the K\"ahler derivative terms,
which we have neglected, are of lower order in the fields than some of the
superpotential terms which we are relying upon. However, as already mentioned in
section~\ref{sec:HigherOrderRB}, the supergravity corrections to the $F$-terms
are indeed negligible because $\W$ is suppressed. We have also checked this
numerically.

\subsection{An example from a heterotic orbifold model}

The aim of this section is to show that the ideas discussed above
can also be applied to string-theoretic models. 
In the following we focus on the models of the `heterotic Mini-Landscape'
\cite{Lebedev:2006kn,Lebedev:2007hv}. These models exhibit the
standard model gauge group and the chiral matter content of the MSSM. They are
based on the \Z6-II orbifold with three factorizable tori (see
\cite{Kobayashi:2004ya,Buchmuller:2006ik} for details). The
discrete symmetry of the geometry leads to a large number of discrete symmetries
governing the couplings of the effective field theory
\cite{Hamidi:1986vh,Dixon:1986qv} (cf.\
also~\cite{Kobayashi:2004ya,Buchmuller:2006ik,Kobayashi:2006wq}).
Apart from various bosonic discrete symmetries, one has a
\begin{equation}\label{eq:DiscreteR}
 [\Z6\times\Z3\times\Z2]_R
\end{equation}
symmetry; other orbifolds have similar discrete symmetries.
Further, in almost all of the Mini-Landscape models there is, at one-loop,
a Fayet-Iliopoulos (FI) $D$-term $\xi$,
\begin{equation}
 V_D~\supset~g^2\,\left(\sum_iq_i\,|\Phi_i|^2+\xi\right)^2\;,
\end{equation}
where the $q_i$ denote the charges under the so-called `anomalous \U1{}'.
It turns out that, in all models with non-vanishing FI term, $\xi$ is of order
$0.1$ (see \cite{Buchmuller:2006ik} for an explicit example).

The first step of our analysis is to identify a set of standard model singlets
$\Phi_i$ with the following properties:
\begin{itemize}
 \item giving VEVs to the $\Phi_i$ allows to cancel the FI
 term;
 \item there is no other field that is singlet under the gauge symmetries left
 unbroken by the $\Phi_i$ VEVs.
\end{itemize}
These properties ensure that the $\Phi_i$ can be consistent with a vanishing
$D$-term potential and that the $F$-terms of all other massless modes vanish,
implying that it is sufficient to derive the superpotential terms involving only
the $\Phi_i$ fields.  

Given non-trivial solutions to the $F$-term equations,
\begin{equation}
 \Phi_i\,\frac{\partial\mathscr{W}}{\partial \Phi_i}~=~0\;,
 \quad\text{with}~\Phi_i\ne0\;,
\end{equation}
one can use complexified gauge transformations to ensure vanishing $D$-terms as
well \cite{Ovrut:1981wa}. Although $D$-term constraints do not fix the scale of
the $\langle \Phi_i\rangle$ in general, the requirement to cancel the FI term
introduces the scale $\sqrt{\xi}\sim 0.3$ into the problem. 
In the following we will search for
solutions of $V_D=V_F=0$ in the regime $|\Phi_i|<1$.
We will explicitly verify that for such solutions the superpotential is
hierarchically small, $\langle\mathscr{W}\rangle\sim \langle \Phi\rangle^N$
where $\langle \Phi\rangle$ denotes the typical size of a VEV. A very important
property of many of these configurations is that all fields acquire
(supersymmetric) masses. Hereby typically only one field -- the would be $R$-axion --  
has a mass of the order $\langle\mathscr{W}\rangle$ 
while the others are much heavier. 

Turning to particular models within the Mini-Landscape we find that the corresponding superpotentials
exhibit accidental $\U1_R$ symmetries that get only broken at rather high
orders $N$. Consequently the analysis becomes very involved, especially when many fields have to be considered. 
This is the case for the phenomenologically very interesting model~1 of \cite{Lebedev:2007hv}, 
where 24 fields have to be switched on.
In this model $\U1_R$ gets broken at order 9, and the superpotential consists of 1816 terms at this order.
To avoid this very complicated setup let us consider another model from the Mini-Landscape which
is easier to handle but nevertheless exhibits the desired features. 

The model we will consider in the following is defined by the gauge shift $V$ and the two Wilson lines $W_{1,2}$,
\begin{eqnarray*}
 V & = & \left(
\begin{array}{llllllll}
 \frac{1}{3} & -\frac{1}{2} & -\frac{1}{2} & 0 & 0 & 0 & 0 & 0 \\
 \frac{1}{2} & -\frac{1}{6} & -\frac{1}{2} & -\frac{1}{2} & -\frac{1}{2} & -\frac{1}{2} & -\frac{1}{2} & \frac{1}{2}
\end{array}
\right)\;,\\
 W_2 & = & \left(
\begin{array}{llllllll}
 \frac{1}{4} & -\frac{1}{4} & -\frac{1}{4} & -\frac{1}{4} & -\frac{1}{4} & \frac{1}{4} & \frac{1}{4} & \frac{1}{4} \\
 1 & -1 & -1 & -1 & -1 & -\frac{1}{2} & -\frac{1}{2} & 2
\end{array}
\right)\;,\\
 W_3 & = & \left(
\begin{array}{llllllll}
 -\frac{1}{2} & -\frac{1}{2} & \frac{1}{6} & \frac{1}{6} & \frac{1}{6} & \frac{1}{6} & \frac{1}{6} & \frac{1}{6} \\
 \frac{1}{3} & 0 & 0 & \frac{2}{3} & \frac{5}{3} & -2 & 2 & 2
\end{array}
\right)\;.
\end{eqnarray*}
The gauge group after compactification is (up to U(1) factors)
\[
[\text{SU}(3) \times  \text{SU}(2)] \times [\text{SU}(3) \times  \text{SU}(2) \times  \text{SU}(2) \times  \text{SU}(2) \times  \text{SU}(2)]\;,
\]
where the two brackets refer to the first and second E$_8$ factor respectively.

We consider the following set of singlet fields with non-vanishing VEVs (in the notation of \cite{Lebedev:2007hv}):
\[
\{\Phi_i\} =\{s_{1},s_{6},s_{7},s_{10},s_{13},s_{14},s_{17},s_{22},s_{24},s_{33}\}\;.
\]
The monomial to cancel the FI term is simply given by $\{s_{33}\}$.
The resulting superpotential exhibits an accidental continuous $R$-symmetry up to order 10 which
is explicitly broken at order 11.
At this order the superpotential consists of 56 terms with 28 independent coefficients,
\begin{eqnarray}
\lefteqn{\W ~=~ \tfrac{1}{288} \lambda_1 s_1 s_{22}^3s_{33}^4  \left(s_{10}^2+s_{17}^2\right)
      +\tfrac{1}{96}  \lambda_2 s_1 s_{22}^2s_{24}s_{33}^4 
	  \left(s_{10}^2+s_{17}^2\right) }\nonumber \\& & {}+ 
       \tfrac{1}{96}  \lambda_3 s_1 s_{22}s_{24}^2s_{33}^4  \left(s_{10}^2+s_{17}^2\right)
      +\tfrac{1}{288} \lambda_4 s_1 s_{24}^3s_{33}^4  \left(s_{10}^2+s_{17}^2\right) \nonumber \\& & {}+ 
       \tfrac{1}{2}   \lambda_5 s_1 s_{22} s_{33}^2  (s_{10}s_6+s_{13} s_{17})
      +\tfrac{1}{2}   \lambda_6 s_1s_{24} s_{33}^2  (s_{10}s_6+s_{13} s_{17}) \nonumber \\& & {}+ 
       \tfrac{1}{144} \lambda_7 s_{22}^3s_{33}^4  (s_{10}s_7+s_{14} s_{17})
      +\tfrac{1}{48}  \lambda_8 s_{22}^2s_{24}s_{33}^4  (s_{10}s_7+s_{14} s_{17}) \nonumber \\& & {}+ 
       \tfrac{1}{48}  \lambda_9 s_{22}s_{24}^2s_{33}^4  (s_{10}s_7+s_{14} s_{17})
      +\tfrac{1}{144} \lambda_{10} s_{24}^3s_{33}^4 (s_{10}s_7+s_{14} s_{17}) \nonumber \\& & {}+ 
       \tfrac{1}{96}  \lambda_{11}s_{10} s_{14} s_{17} s_{22}s_{33}^4 s_7  (s_{10}s_7+s_{14} s_{17})
      +\tfrac{1}{96}  \lambda_{12}s_{10} s_{14} s_{17} s_{24}s_{33}^4s_7  (s_{10}s_7+s_{14} s_{17}) \nonumber \\& & {}+ 
       \tfrac{1}{288} \lambda_{13}s_{14} s_{22}s_{33}^4s_7  \left(s_{13}^3s_7+s_{14} s_6^3\right)
      +\tfrac{1}{288} \lambda_{14}s_{14}s_{24}s_{33}^4s_7  \left(s_{13}^3  s_7+s_{14}s_6^3\right) \nonumber \\& & {}+ 
       \tfrac{1}{288} \lambda_{15}s_{14} s_{22} s_{33}^4s_7  \left(s_{10}^3 s_{14} +s_{17}^3s_7\right)
      +\tfrac{1}{288} \lambda_{16}s_{14}s_{24}s_{33}^4s_7  \left(s_{10}^3 s_{14}+s_{17}^3 s_7\right) \nonumber \\& & {}+ 
       \tfrac{1}{2}   \lambda_{17} s_{22}s_{33}^2  (s_{13} s_{14}+s_6s_7)
      +\tfrac{1}{2}   \lambda_{18}s_{24}s_{33}^2  (s_{13} s_{14}+s_6s_7) \nonumber \\& & {}+ 
       \tfrac{1}{96}  \lambda_{19}s_{13} s_{14} s_{22}s_{33}^4s_6s_7  (s_{13} s_{14}+s_6s_7)
      +\tfrac{1}{96}  \lambda_{20}s_{13} s_{14}s_{24}s_{33}^4s_6s_7  (s_{13} s_{14}+s_6 s_7) \nonumber \\& & {}+ 
       \tfrac{1}{864} \lambda_{21}s_{22}s_{33}^4  \left(s_{10}^3s_7^3+s_{14}^3 s_{17}^3\right)
      +\tfrac{1}{864} \lambda_{22}s_{24}s_{33}^4  \left(s_{10}^3s_7^3+s_{14}^3 s_{17}^3\right) \nonumber \\& & {}+ 
       \tfrac{1}{288} \lambda_{23}s_{13} s_{22}s_{33}^4s_6   \left(s_{13}s_7^3+s_{14}^3s_6\right)
      +\tfrac{1}{288} \lambda_{24} s_{13}s_{24}s_{33}^4s_6    \left(s_{13}s_7^3+s_{14}^3s_6\right) \nonumber \\& & {}+ 
       \tfrac{1}{288} \lambda_{25} s_{10} s_{17} s_{22}s_{33}^4  \left(s_{10} s_{14}^3+s_{17}s_7^3\right)
      +\tfrac{1}{288} \lambda_{26} s_{10} s_{17}s_{24}s_{33}^4  \left(s_{10} s_{14}^3+s_{17}s_7^3\right)  \nonumber \\& & {}+  
       \tfrac{1}{864} \lambda_{27} s_{22}s_{33}^4  \left(s_{13}^3 s_{14}^3+s_6^3s_7^3\right)
      +\tfrac{1}{864} \lambda_{28} s_{24}s_{33}^4  \left(s_{13}^3 s_{14}^3+s_6^3s_7^3\right) \;.
	  \label{eq:W80}
\end{eqnarray}
If all these fields acquire VEVs, two U(1) factors get broken, 
one of them corresponding to the anomalous U(1).
The charges of the ten singlets with respect to the two broken U(1) factors are
\begin{align}
q_\text{anom}^i &= (6, 11, 17, 11, 11, 17, 11, 28, 28, -28) \;, \nonumber \\ 
q_{\U1}^i &= (-2, 1, -1, 1, 1, -1, 1, 0, 0, 0) \;.
\end{align}

The full quantum numbers of the $s_i$ are (in the conventions of
\cite{Lebedev:2007hv})
\begin{footnotesize}
\begin{center}
\begin{tabular}{|c|c|c|c|c|c|c|c|c|c|c|c|c|c|c|c|c|}
\hline
 field & $k$ & $n_3$ & $n_2$ & $n_2'$ & $q_\gamma$ & $R_1$ & $R_2$ & $R_3$ & irrep &
 $q_Y$ & $q_2$ & $q_3$ & $q_4$ & $q_5$ & $q_6$ & $q_7$ \\
\hline 
 $s_{1}$ & $0$ & $*$ & $*$ & $*$ & $0$ & $-1$ & $0$ & $0$ & $\left(\boldsymbol{1},\boldsymbol{1};\boldsymbol{1},\boldsymbol{1},\boldsymbol{1},\boldsymbol{1},\boldsymbol{1}\right)$ & $0$ & $-1$ & $-1$ & $0$ & $0$ & $0$ & $0$ \\
 $s_{6}$ & $1$ & $0$ & $0$ & $0$ & $0$ & $-\frac{1}{6}$ & $\frac{2}{3}$ & $-\frac{1}{2}$ & $\left(\boldsymbol{1},\boldsymbol{1};\boldsymbol{1},\boldsymbol{1},\boldsymbol{1},\boldsymbol{1},\boldsymbol{1}\right)$ & $0$ & $\frac{1}{3}$ & $\frac{1}{2}$ & $\frac{1}{2}$ & $0$ & $\frac{1}{3}$ & $0$\\
 $s_{7}$ & $1$ & $0$ & $0$ & $0$ & $0$ & $\frac{5}{6}$ & $-\frac{1}{3}$ & $-\frac{1}{2}$ & $\left(\boldsymbol{1},\boldsymbol{1};\boldsymbol{1},\boldsymbol{1},\boldsymbol{1},\boldsymbol{1},\boldsymbol{1}\right)$ & $0$ & $-\frac{2}{3}$ & $-\frac{1}{2}$ & $\frac{1}{2}$ & $0$ & $\frac{1}{3}$ & $0$\\
 $s_{10}$ & $1$ & $0$ & $0$ & $0$ & $0$ & $\frac{11}{6}$ & $-\frac{1}{3}$ & $-\frac{1}{2}$ & $\left(\boldsymbol{1},\boldsymbol{1};\boldsymbol{1},\boldsymbol{1},\boldsymbol{1},\boldsymbol{1},\boldsymbol{1}\right)$ & $0$ & $\frac{1}{3}$ & $\frac{1}{2}$ & $\frac{1}{2}$ & $0$ & $\frac{1}{3}$ & $0$\\
 $s_{13}$ & $1$ & $0$ & $0$ & $1$ & $0$ & $-\frac{1}{6}$ & $\frac{2}{3}$ & $-\frac{1}{2}$ & $\left(\boldsymbol{1},\boldsymbol{1};\boldsymbol{1},\boldsymbol{1},\boldsymbol{1},\boldsymbol{1},\boldsymbol{1}\right)$ & $0$ & $\frac{1}{3}$ & $\frac{1}{2}$ & $\frac{1}{2}$ & $0$ & $\frac{1}{3}$ & $0$\\
 $s_{14}$ & $1$ & $0$ & $0$ & $1$ & $0$ & $\frac{5}{6}$ & $-\frac{1}{3}$ & $-\frac{1}{2}$ & $\left(\boldsymbol{1},\boldsymbol{1};\boldsymbol{1},\boldsymbol{1},\boldsymbol{1},\boldsymbol{1},\boldsymbol{1}\right)$ & $0$ & $-\frac{2}{3}$ & $-\frac{1}{2}$ & $\frac{1}{2}$ & $0$ & $\frac{1}{3}$ & $0$\\
 $s_{17}$ & $1$ & $0$ & $0$ & $1$ & $0$ & $\frac{11}{6}$ & $-\frac{1}{3}$ & $-\frac{1}{2}$ & $\left(\boldsymbol{1},\boldsymbol{1};\boldsymbol{1},\boldsymbol{1},\boldsymbol{1},\boldsymbol{1},\boldsymbol{1}\right)$ & $0$ & $\frac{1}{3}$ & $\frac{1}{2}$ & $\frac{1}{2}$ & $0$ & $\frac{1}{3}$ & $0$\\
 $s_{22}$ & $2$ & $0$ & $*$ & $*$ & $0$ & $-\frac{1}{3}$ & $-\frac{2}{3}$ & $0$ & $\left(\boldsymbol{1},\boldsymbol{1};\boldsymbol{1},\boldsymbol{1},\boldsymbol{1},\boldsymbol{1},\boldsymbol{1}\right)$ & $0$ & $-\frac{1}{3}$ & $0$ & $1$ & $0$ & $\frac{2}{3}$ & $0$\\
 $s_{24}$ & $2$ & $0$ & $*$ & $*$ & $1$ & $-\frac{1}{3}$ & $-\frac{2}{3}$ & $0$ & $\left(\boldsymbol{1},\boldsymbol{1};\boldsymbol{1},\boldsymbol{1},\boldsymbol{1},\boldsymbol{1},\boldsymbol{1}\right)$ & $0$ & $-\frac{1}{3}$ & $0$ & $1$ & $0$ & $\frac{2}{3}$ & $0$\\
 $s_{33}$ & $4$ & $0$ & $*$ & $*$ & $\frac{1}{2}$ & $-\frac{2}{3}$ & $-\frac{1}{3}$ & $0$ & $\left(\boldsymbol{1},\boldsymbol{1};\boldsymbol{1},\boldsymbol{1},\boldsymbol{1},\boldsymbol{1},\boldsymbol{1}\right)$ & $0$ & $\frac{1}{3}$ & $0$ & $-1$ & $0$ & $-\frac{2}{3}$ & $0$\\
\hline
\end{tabular}
\end{center}
\end{footnotesize}
The structure of the superpotential \eqref{eq:W80} is governed by a $D_4$
symmetry,\footnote{$\mathrm{D}_4\subset$ O$(2)$ is the dihedral group of order
$8$, the symmetry group of a square.} under which 6 fields combine to 3
doublets,
\[
   d_1~=~\left(\begin{array}{c}s_6\\ s_{13}\end{array}\right)
   \;,\quad
   d_2~=~\left(\begin{array}{c}s_7\\ s_{14}\end{array}\right)	
   \quad\text{and}\quad
   d_3~=~\left(\begin{array}{c}s_{10}\\ s_{17}\end{array}\right)
   \;.	
\] 
It further turns out that any allowed term (at order less or equal to 11)
involves at least two such doublets. Another observation is that $s_{33}$
appears only with even powers.

The solutions of the global $F$-term equations depend on the precise values of
the $\lambda_i$ coefficients. As we do not yet know how to compute these
coefficients, we can only argue that point-like solutions to the $F$- and
$D$-term equations exist. A consistency check for this assertion is as follows:
rather than solving for the fields we solve the $F$-term equations for the 28
coefficients $\lambda_i$, setting the fields to `desirable' values consistently
with $D$-flatness. 
An example for a, thus obtained, `vacuum configuration' is given by 
\begin{eqnarray}
\langle s_{1} \rangle &=& - \frac{1}{10}\;,\; \langle s_{6} \rangle ~=~  \frac{1}{10}\;,\; 
\langle s_{7} \rangle ~=~  \frac{1}{10}\;,\;\langle s_{10} \rangle ~=~  \frac{1}{10}\;,\;
\langle s_{13} \rangle ~=~  \frac{1}{10}\;,\;
\nonumber \\
\langle s_{14} \rangle &=&  \frac{1}{10}\;,\;
\langle s_{17} \rangle ~=~  \frac{1}{10}\;,\;\langle s_{22} \rangle ~=~  \frac{1}{10}\;,\;
\langle s_{24} \rangle ~=~  \frac{1}{10}\;,\;\langle s_{33} \rangle ~=~  -\frac{\sqrt{1401}}{20\sqrt{70}}
\nonumber
\end{eqnarray}
with  all $\lambda_i=0.01$ except for
\begin{eqnarray}
 \lambda_1 &\sim& 0.482 \;,\; \lambda_5 ~\sim~ -0.01 \;,\; \lambda_{13} ~\sim~  -0.29 \;,\; \lambda_{14} ~\sim~  -0.22 \;,\; 
\nonumber \\
\lambda_{17} &\sim & -0.001 \;,\;  \lambda_{8} ~\sim~  0.001 \;,\;  \lambda_{9} ~\sim~  0.001
\;,\; \lambda_{18} ~\sim~  0.001 \nonumber\;.
\end{eqnarray}
The coefficients $(\lambda_8,\lambda_9,\lambda_{18})$ are conveniently chosen
and $(\lambda_1, \lambda_5,  \lambda_{13}, \lambda_{14}, \lambda_{17})$  are
fixed in terms of the remaining coefficients. 
The resulting vacuum expectation value for $\W$ in units of $M_\text{P}$ is given by
\[
 \langle \W \rangle ~\sim~ 1.1 \cdot 10^{-12}
\]
while the mass eigenvalues resulting from the superpotential read
\begin{eqnarray}
(m_i) & =& \left(1\cdot10^{-5},\;1\cdot10^{-5},\; 1.4 \cdot10^{-9},\; 
 5.1 \cdot10^{-10},\; 2.7 \cdot10^{-10},\; \right. \nonumber \\
 && \; \; \left. 2.2 \cdot10^{-10},\; 1\cdot10^{-10},\; 3\cdot10^{-11},\; 0,\; 0
 \right)\;.
\end{eqnarray}
The two massless fields obtain masses from the $D$-term potentials corresponding to the 
two broken U(1) factors and the absorption of the Goldstone modes, respectively.
We see that the lightest mass eigenstate is of order $\langle \Phi \rangle^{N-2} \sim 10^2 \langle \W \rangle$
as expected.

Altogether we have argued that in the model under consideration one may obtain
isolated supersymmetric field configurations with $|\Phi_i|<1$ where the
VEV of the perturbative superpotential $\langle\mathscr{W}\rangle$ is 
hierarchically small. It appears highly desirable to rigorously prove that such
configurations exist in explicit string-derived models. This, however, requires
knowledge of the coefficients $\lambda_i$, which is not yet available.

\section{Conclusions}
\label{sec:Conclusions}

We have investigated the mechanism of generating a hierarchically
small superpotential expectation value $\vev\W$ by an approximate $R$-symmetry. 
We have recapitulated that, in the presence of such a symmetry, $\vev\W$ can be 
highly suppressed if the typical scalar expectation values are only somewhat below 
the fundamental scale~\cite{Kappl:2008ie}. In the limiting case of an exact $R$-symmetry, we 
showed that there exist examples of generic models where $R$-symmetry is broken 
spontaneously in a supersymmetric vacuum. By adding higher-order polynomial terms 
to the superpotential which break $R$-symmetry explicitly, one may then construct 
vacua with $\vev\W$ given by high powers of small field expectation values. If 
they can be uplifted to Minkowski minima of the scalar potential, one obtains 
potentially realistic vacua with naturally small gravitino mass. 

The main point of this analysis is the observation that an (approximate)
$R$-symmetry not only allows us to control $\vev\W$, but also the MSSM $\mu$
term, if the Higgs fields $H_u$ and $H_d$ are singlets with respect to all
symmetries and in particular have trivial $R$-charges. For models with generic 
superpotential coefficients, we proved that $\mu\sim\vev\W$ in Planck units, 
i.e.\ $\mu\sim m_{3/2}$.

We have also commented on the situation in string-derived models. We analyzed
scenarios in which an $R$-symmetry and gauge invariance in higher dimensions
relate the $\mu$ term to $\vev\W$. We have identified explicit string-derived
models with the chiral spectrum of the MSSM in which this analysis can be
applied. These models indeed exhibit approximate $\U1_R$ symmetries, deriving
from high-power discrete  symmetries, which can explain a highly suppressed
$\vev\W$. We further discussed  explicit examples in which such suppressed
$\vev\W$ emerge, while all fields are stabilized. To rigorously prove that such
configurations exist in string-derived models, and to study their phenomenology,
however, will require a detailed understanding of the coupling strengths. 

The main focus of this analysis was the role of approximate $R$-symmetries in
generating a suppressed $\vev\W$ and a $\mu$ term of a similar size. However,
in  our examples the origin of supersymmetry breakdown remains obscure. To be
able  to relate $\mu$ to the MSSM soft masses properly, a better understanding
of the mechanism of supersymmetry breakdown (or the so-called ``uplifting'') is
required.

\bigskip\bigskip

\noindent
\textbf{Acknowledgments.} We would like to thank  Fawzi Boudjema, Laura Covi,
Arthur Hebecker and Joerg Jaeckel for useful discussions. This research was
supported by the cluster of excellence Origin and Structure of the Universe and
the \mbox{SFB-Transregio}  27 "Neutrinos and Beyond" by Deutsche
Forschungsgemeinschaft (DFG). M.R.\ would like to thank the Aspen
Center for Physics, where some of this work has been carried out, for
hospitality and support. F.B.\ would like to thank LPSC Grenoble for hospitality
and support.

\appendix

\section[Identifying accidental $R$-symmetries]{%
Identifying accidental $\boldsymbol{R}$-symmetries}
\label{app:AccidentalRSymmetries}

Consider the superpotential of some fields $\Phi_i$. Truncate it at a certain
order, such that it is a sum of monomials of degree $\le D$,
\begin{equation}\label{eq:WR}
 \W~=~\sum_{m=1}^M \Phi_1^{\nu_1^{(m)}}\cdots \Phi_N^{\nu_N^{(m)}}\;,
\end{equation}
where $\nu_i^{(m)}\in\mathbbm{N}_0$. The exponent vectors
$\nu^{(m)}=(\nu_1^{(m)},\dots \nu_N^{(m)})$ can be used for the task of
identifying accidental $R$-symmetries. Define a matrix
\begin{equation}
 A~=~\left(\begin{array}{ccc}
 \nu_1^{(1)} & \dots & \nu_N^{(1)}\\
 \vdots & & \vdots\\
 \nu_1^{(M)} & \dots & \nu_N^{(M)} 
 \end{array}\right)\;,
\end{equation}
with $M$ denoting the number of monomials appearing in \eqref{eq:WR}.
An (accidental) \U1 symmetry exists only if the equation
\begin{equation}\label{eq:AccU1}
 A\cdot x~=~0
\end{equation}
possesses a non-trivial solution. The entries of such a solution are the charges
w.r.t.\ the (accidental) \U1.

In order to identify (accidental) $R$-symmetries, one has to solve the
equation
\begin{equation}
 A\cdot x_R~=~\left(\begin{array}{c}2\\ \vdots\\ 2\end{array}\right)\;.
\end{equation}
The entries of $x_R$ denote the $R$-charges. It is clear that, given a solution
$x_R$, one can always add a `bosonic' solution of \eqref{eq:AccU1} to obtain
`another' $R$-symmetry. However, as the superpotential is a gauge invariant
quantity, this does not change any of the conclusions presented in the main
text.

\bibliography{Orbifold}

\providecommand{\bysame}{\leavevmode\hbox to3em{\hrulefill}\thinspace}
\begin{thebibliography}{10}

\bibitem{'tHooft:1979bh}
G.~'t~Hooft, NATO Adv. Study Inst. Ser. B Phys. \textbf{59} (1980), 135.

\bibitem{Witten:1981nf}
E.~Witten, Nucl. Phys. \textbf{B188} (1981), 513.

\bibitem{Kappl:2008ie}
R.~Kappl, H.~P. Nilles, S.~Ramos-S{\'a}nchez, M.~Ratz, K.~Schmidt-Hoberg, and
  P.~K. Vaudrevange, Phys. Rev. Lett. \textbf{102} (2009), 121602,
  [0812.2120].

\bibitem{Kim:1983dt}
J.~E. Kim and H.~P. Nilles, Phys. Lett. \textbf{B138} (1984), 150.

\bibitem{Giudice:1988yz}
G.~F. Giudice and A.~Masiero, Phys. Lett. \textbf{B206} (1988), 480--484.

\bibitem{Buchmuller:2006ik}
W.~Buchm{\"u}ller, K.~Hamaguchi, O.~Lebedev, and M.~Ratz, Nucl. Phys.
  \textbf{B785} (2007), 149--209,  [hep-th/0606187].

\bibitem{Luty:1995sd}
M.~A. Luty and W.~Taylor, Phys. Rev. \textbf{D53} (1996), 3399--3405,
  [hep-th/9506098].

\bibitem{Froggatt:1978nt}
C.~D. Froggatt and H.~B. Nielsen, Nucl. Phys. \textbf{B147} (1979), 277.

\bibitem{Nelson:1993nf}
A.~E. Nelson and N.~Seiberg, Nucl. Phys. \textbf{B416} (1994), 46--62,
  [hep-ph/9309299].

\bibitem{Ray:2007wq}
S.~Ray,  (2007),  0708.2200.

\bibitem{GomezReino:2006dk}
M.~Gomez-Reino and C.~A. Scrucca, JHEP \textbf{05} (2006), 015,
  [hep-th/0602246].

\bibitem{Lebedev:2006qq}
O.~Lebedev, H.~P. Nilles, and M.~Ratz, Phys. Lett. \textbf{B636} (2006),
  126--131,  [hep-th/0603047].

\bibitem{Brummer:2006dg}
F.~Br{\"u}mmer, A.~Hebecker, and M.~Trapletti, Nucl. Phys. \textbf{B755}
  (2006), 186--198,  [hep-th/0605232].

\bibitem{Weinberg:1988cp}
S.~Weinberg, Rev. Mod. Phys. \textbf{61} (1989), 1--23.

\bibitem{Casas:1992mk}
J.~A. Casas and C.~Mu{\~n}oz, Phys. Lett. \textbf{B306} (1993), 288--294,
  [hep-ph/9302227].

\bibitem{Kreuzer:1992bi}
M.~Kreuzer and H.~Skarke, Commun. Math. Phys. \textbf{150} (1992), 137,
  [hep-th/9202039].

\bibitem{Lebedev:2007hv}
O.~Lebedev, H.~P. Nilles, S.~Raby, S.~Ramos-S{\'a}nchez, M.~Ratz, P.~K.~S.
  Vaudrevange, and A.~Wingerter, Phys. Rev. \textbf{D77} (2007), 046013,
  [arXiv:0708.2691 [hep-th]].

\bibitem{Antoniadis:1994hg}
I.~Antoniadis, E.~Gava, K.~S. Narain, and T.~R. Taylor, Nucl. Phys.
  \textbf{B432} (1994), 187--204,  [hep-th/9405024].

\bibitem{Brignole:1996xb}
A.~Brignole, L.~E. Ib{\'a}{\~n}ez, and C.~Mu{\~n}oz, Phys. Lett. \textbf{B387}
  (1996), 769--774,  [hep-ph/9607405].

\bibitem{Lebedev:2006kn}
O.~Lebedev, H.~P. Nilles, S.~Raby, S.~Ramos-S{\'a}nchez, M.~Ratz, P.~K.~S.
  Vaudrevange, and A.~Wingerter, Phys. Lett. \textbf{B645} (2007), 88,
  [hep-th/0611095].

\bibitem{Cvetic:1988yw}
M.~Cveti{\v{c}}, J.~Louis, and B.~A. Ovrut, Phys. Lett. \textbf{B206} (1988),
  227.

\bibitem{Brignole:1997dp}
A.~Brignole, L.~E. Ib{\'a}{\~n}ez, and C.~Mu{\~n}oz,  (1997),  hep-ph/9707209,
  in Kane, G.L. (ed.): Perspectives on supersymmetry, 125-148.

\bibitem{Dixon:1986qv}
L.~J. Dixon, D.~Friedan, E.~J. Martinec, and S.~H. Shenker, Nucl. Phys.
  \textbf{B282} (1987), 13--73.

\bibitem{Blaszczyk:2009in}
M.~Blaszczyk et~al., Phys. Lett. \textbf{B683} (2010), 340--348,  [0911.4905].

\bibitem{Dundee:2010sb}
B.~Dundee, S.~Raby, and A.~Westphal,  (2010),  1002.1081.

\bibitem{Choi:2003kq}
K.-w. Choi et~al., JHEP \textbf{02} (2004), 037,  [hep-ph/0312178].

\bibitem{Brummer:2009ug}
F.~Br{\"u}mmer, S.~Fichet, A.~Hebecker, and S.~Kraml, JHEP \textbf{08} (2009),
  011,  [0906.2957].

\bibitem{Burdman:2002se}
G.~Burdman and Y.~Nomura, Nucl. Phys. \textbf{B656} (2003), 3--22,
  [hep-ph/0210257].

\bibitem{Hosteins:2009xk}
P.~Hosteins, R.~Kappl, M.~Ratz, and K.~Schmidt-Hoberg, JHEP \textbf{07} (2009),
  029,  [0905.3323].

\bibitem{Lee:2003mc}
H.~M. Lee, H.~P. Nilles, and M.~Zucker, Nucl. Phys. \textbf{B680} (2004),
  177--198,  [hep-th/0309195].

\bibitem{Buchmuller:2005jr}
W.~Buchm{\"u}ller, K.~Hamaguchi, O.~Lebedev, and M.~Ratz, Phys. Rev. Lett.
  \textbf{96} (2006), 121602,  [hep-ph/0511035].

\bibitem{Kobayashi:2004ya}
T.~Kobayashi, S.~Raby, and R.-J. Zhang, Nucl. Phys. \textbf{B704} (2005),
  3--55,  [hep-ph/0409098].

\bibitem{Hamidi:1986vh}
S.~Hamidi and C.~Vafa, Nucl. Phys. \textbf{B279} (1987), 465.

\bibitem{Kobayashi:2006wq}
T.~Kobayashi, H.~P. Nilles, F.~Pl{\"o}ger, S.~Raby, and M.~Ratz, Nucl. Phys.
  \textbf{B768} (2007), 135--156,  [hep-ph/0611020].

\bibitem{Ovrut:1981wa}
B.~A. Ovrut and J.~Wess, Phys. Rev. \textbf{D25} (1982), 409.

\end{thebibliography}
\bibliographystyle{ArXiv}

\end{document}